\documentclass[a4 paper, 12 pt] {article}
\usepackage{amsmath,amssymb}
\def\be{\begin{equation}}
\def\ee{\end{equation}}
\def\bdi{\begin{displaymath}}
\def\edi{\end{displaymath}}
\def\br{\begin{eqnarray}}
\def\er{\end{eqnarray}}

\def\RR{{\rm I\kern-.1567em R}}                              
 \def\CC{{\rm C\kern-4.7pt                                    
 \vrule height 7.7pt width 0.4pt depth -0.5pt \phantom {.}}} 
 \def\ZZ{{\sf Z\kern-4.5pt Z}}                                

\begin{document}

\begin{titlepage}
\vspace*{-3 cm}
\noindent

\vskip 2cm
\begin{center}
{\Large\bf 
Compact self-gravitating solutions of quartic (K)
fields in brane cosmology}
\vglue 1  true cm

C. Adam$^{a*}$,   N. Grandi$^{\, b**}$, P. Klimas$^{a***}$,
J. S\'anchez-Guill\'en$^{a\dagger}$,
and A. Wereszczy\'nski$^{c\dagger\dagger}$
\vspace{1 cm}

\small{ $^{a)}$Departamento de Fisica de Particulas, Universidad
     de Santiago}
     \\
     \small{ and Instituto Galego de Fisica de Altas Enerxias (IGFAE)}
     \\ \small{E-15782 Santiago de Compostela, Spain}
     \\ \small{ $^b)$ IFLP-CONICET }
     \\ \small{cc67 CP1900, La Plata, Argentina}
      \\ \small{ $^{c)}$Institute of Physics,  Jagiellonian
     University,}
     \\ \small{ Reymonta 4, 30-059 Krak\'{o}w, Poland}

\medskip
\end{center}

\normalsize
\vskip 0.2cm

\begin{abstract}
Recently we proposed that 
{\em K} fields, that is, fields with a non-standard kinetic term, 
may provide a mechanism for the generation of thick branes,
based on the following observations. Firstly, {\em K} field theories  
allow for
soliton solutions with compact support, i.e., compactons. Compactons
in 1+1 dimensions may give rise to topological defects of the domain
wall type and with finite thickness in higher dimensions. Secondly,
propagation of linear perturbations is  confined  inside the compacton
domain wall. Further, these linear perturbations inside the topological defect 
are of the standard type, in spite of
the non-standard kinetic term. 
Thirdly, when gravity is taken into
account, location of gravity in the sense of Randall--Sundrum works for
these compacton domain walls provided that the backreaction of gravity
does not destabilize the compacton domain wall.
\\
It is the purpose of the present 
paper to investigate in detail the existence and stability of 
compacton domain walls in the full {\em K} field and gravity system, using
both analytical and numerical methods.
We find that the existence of the domain wall in the full system 
requires a 
correlation  between the gravitational constant and the bulk cosmological
constant.

\end{abstract}

\vfill

{\footnotesize
$^*$adam@fpaxp1.usc.es

$^{**}$grandi@fisica.unlp.edu.ar

$^{***}$klimas.ftg@gmail.com

$^{\dagger}$joaquin@fpaxp1.usc.es

$^{\dagger\dagger}$wereszczynski@th.if.uj.edu.pl }

\end{titlepage}

\section{Introduction}
Recently, the idea that the visible universe with its 3+1 dimensions is
embedded into some higher dimensional space has received considerable 
attention. 
Consistency with observations requires that the propagation of matter and
fields is restricted to this 3+1 dimensional subspace, at least for not too
high energies. In this context, two slightly different scenarios have been
studied.
The subspace may either be of the topological defect type, in which case it
has a finite, although probably very small, extension in the additional
dimensions. Some first proposals of this type have already been made in the
eighties, see \cite{Akama1}, \cite{RuSh1}. In the last years, interest in this
type of ``topological defect'' universes has increased significantly, and the
name of ``thick branes'' has become customary for these objects. Some recent
work may be found, e.g., in \cite{CEHS1} - \cite{AnVe1}.
The other possibility is that the subspace is strictly 3+1 dimensional,
in which case it is known as a ``three-brane''. Investigation of these 
three-branes started in the nineties, and the literature on this subject is
too numerous to be quoted here.  
Some recent reviews, in which also further references may be found, are, 
e.g.,  \cite{Lan1} - \cite{BrBr1}.

It is one of the crucial features of a ``thick brane'' cosmological model
that a dynamical
mechanism must exist which provides the
confinement of all matter fields to the subspace. 
We recently proposed such a mechanism \cite{AGSGW1}
which uses as its main
ingredients a scalar field theory 
with a non-standard kinetic term ({\em K} field theory) and the
observation that topological defects with a compact support
(compactons) exist in this theory \cite{ASGW1}.
For different aspects of {\em K} field theory, we refer, e.g., to
\cite{APDM1} - \cite{bazeia}, whereas the theory of compactons is discussed, 
e.g., in \cite{Arodz1} - \cite{Arodz3}.
Interestingly, the complete suppression of the
propagation of fields outside the support of the compacton is an
automatic result of the model proposed in \cite{AGSGW1}. 
Further, the propagation of linear perturbations inside the
topological defect (i.e., inside the brane) is standard, in close similarity
to the Kaluza--Klein reduction, in spite
of the non-standard kinetic term. Specifically, there are no
tachyons on the brane, and the evolution of linear perturbations
is both unitary and causal. Inside the brane, the only remaining
effect of the original {\em K} field theory resides in the values of the
masses of the (Klein--Gordon type) linear fluctuation field.

For the full theory, with gravitation coupled minimally to the scalar field,
we showed in the same paper that, provided the compacton domain
wall is not destabilized by the addition of gravity, bulk gravity
solutions of the Randall--Sundrum type \cite{RaSu1}
(that is, localization of
gravity on the brane) do exist. However, the analysis of
the stability of the full 
{\em K} field plus gravity system and specifically of the existence of the
compacton domain walls in the full system was not done in
Ref. \cite{AGSGW1}.
It is the purpose of the present paper to perform precisely this analysis.

In Section 2, we briefly present the model together with the resulting 
field equations and their vacuum solution. In Section 3 we present some
analytic investigations of the system without gravity for later convenience.
In Section 4 we turn to the analytic investigation of the full system with
gravitation included. We perform a power series expansion about the boundary
of the domain wall as well as about its center. Further, we establish some
global properties of the resulting system of nonlinear evolution equations.
The result of this analysis is that the compacton domain wall does not exist 
for completely generic values of the two remaining parameters (the
gravitational coupling $\kappa$ and the bulk cosmological constant
$\Lambda$). Instead, the existence of the compacton requires a correlation
between these two parameters, 
that is, for a given $\kappa$ the compacton exists if $\Lambda$
takes a fixed $\kappa$ dependent value. In different words, the space of
compacton solutions forms a one-dimensional curve in the $\kappa$-$\Lambda$
plane. In Section 5 we turn to a numerical invastigation of the full
system. We determine the correlation between $\kappa$  and $\Lambda$
(i.e., the above mentioned line in the  $\kappa$-$\Lambda$ plane) to a high
precision. Concretely, we fine-tune the value of $\kappa$ for a given
$\Lambda$. 
Further we determine the compacton radius, that is, the extension
of the domain wall in the transverse direction. In addition, we display 
figures providing some details of the numerical investigation.   In Section
6 we study the issue of stability of the compact domain walls established
in the previous sections under linear fluctuations.  Specifically, we prove
by a combination of analytical and numerical methods that the domain wall
is stable under fluctuations of the scalar {\em K} 
field also in the full theory,
with gravitational backreaction included. Section 7 contains a discussion of
our results.

\section{The model}

The action is (for details we refer to \cite{AGSGW1})
\begin{equation}
  S=\int d^5x \sqrt {|g|}\left( \kappa^{-2}(R - \Lambda )+ 
4 |X|X - V(\xi)\right)
\end{equation}
where $\Lambda$ is the cosmological constant ,
and $X$ now includes the metric 
\begin{equation}
  X = \frac 12 g^{MN}\partial_M \xi\partial_N\xi .
\end{equation}
Further, the potential is
\be  
V(\xi ) \equiv
3 \lambda^4 (\xi^2 - a^2)^2 , 
\ee
We will choose a 5D metric ansatz with a Minkowskian 4D slice, written in 
the form
\begin{equation}
  ds^2=e^{-A(y)}\left(dt^2-d\vec x^{\,2}\right)-dy^2 .
\end{equation}
We remind the reader that - up to coordinate transformations - this metric
ansatz is the most general one 
having the Minkowski space symmetries ISO(3,1) on the brane, see e.g.
\cite{CEHS1}. 
For the scalar field we assume $\xi=\xi(y)$.
Then the field equation for $\xi$ is
\begin{equation} \label{Eqn-y-1}
-8 A_y \xi^3_y + 12 \xi^2_y \xi_{yy} = 12 \lambda^4 \xi (\xi^2 -a^2)
\end{equation}
and the independent components of the Einstein equations read
\begin{eqnarray} \label{Eqn-y-2}
&&\frac34 A_{yy} -  {A_y^{\, 2}} =
\frac13\left[ \Lambda  + \kappa^2 3 \lambda^4 (\xi^2 -a^2)^2 \right]
 \\
&&\frac34A_{yy} = \kappa^2\xi_y^{\, 4} .
\label{Eqn-y-3}
\end{eqnarray}
At first sight one might think that the system is overdetermined 
because there are three ODEs for the two unknown
functions $\xi$ and $A$, but
the field equation for $\xi$ is not
independent of the Einstein equations.
Indeed, when one calculates the $y$ derivative of Eq. (\ref{Eqn-y-2}) and
replaces $A_{yy}$ and $A_{yyy}$ with the help of Eq. (\ref{Eqn-y-3}), one
recovers  Eq. (\ref{Eqn-y-1}). 

Remark: the derivation of the field equation for $\xi$ from the Einstein
equations requires a division by $\kappa^2$ and another division by $\xi_y$,
therefore the field equation is a consequence of the Einstein equations
provided that these two quantities are nonzero. In the vacuum sector, where
$\xi_y =0$, the field equation for $\xi $ is more restrictive and requires
that $\xi =\pm a$, whereas the Einstein equations would allow for more general
constant values of $\xi$.

For later convenience, we rewrite the above system of equations in
dimensionless quantities, by rescaling $\xi \to a \xi$, $V \to a^4 V$,
and $z=\lambda y$, $\lambda^{-1}
\Lambda \to \Lambda $, $\lambda \kappa \to \kappa $. Further, we introduce
the derivative $C(z) =A_z (z)$ as field variable, and get
\br \label{eq-0-g-z}
-\frac23 C \xi^3_z + \xi^2_z \xi_{zz} &=& \xi (\xi^2 -1) \\   \label{eq-1-g-z}
\frac34 C_{z} - C^2 &=& \frac13 \Lambda  + \kappa^2 (\xi^2 -1)^2 \\
\label{eq-2-g-z}
\frac34 C_{z} &=& \kappa^2 \xi_z^4 .
\er
This system still has the vacuum solution 
\be
\xi =\pm 1 = {\rm const. }\quad ,  \qquad A= 
\sqrt{ \frac{\bar \Lambda}{ 3}}\,|z|+ {\rm const. }  \quad \Rightarrow
\quad C=  \sqrt{ \frac{\bar \Lambda}{ 3}} {\rm sign}(z)
\ee
where 
\be
\bar \Lambda \equiv - \Lambda \ge 0
\ee
is a positive constant.

\section{Non-gravitational case}

The system without gravitation has the field equation
\be \label{eq-no-g-z}
\xi^2_z \xi_{zz} = \xi (\xi^2 -1)
\ee
with the compacton solution
\begin{equation}
\xi (z) = \left\{
\begin{array}{lc}
- 1 & \quad (z - z_0) \leq - \frac{\pi}{2 }  \\
 \sin  z & \quad
-\frac{\pi}{2 } \leq (z-z_0) \leq
\frac{\pi}{2 }  \\
1 & \quad (z -z_0) \geq \frac{\pi}{2  },
\end{array}
\right. \label{compacton sol}
\end{equation}
Here $z_0$ is the compacton center, which we shall frequently choose equal to
$z_0=0$. Further, the compacton reaches the vacuum values $\xi =\pm 1$ at
$z_{\pm } = z_0 \pm \frac{\pi}{2}$. We will use the notation $z_0$, $z_\pm$ 
for the center and boundary of the compacton, respectively, also in the case
with gravitation.

Although we know the explicit solution in the case without gravity, we want to
study the behaviour of the solution near the lower boundary and the
center for later
convenience. We assume $z_0 =0$ for the moment and introduce the new variable
$t= z-z_- \equiv z+\frac{\pi}{2}$ (where $\xi (t=0)  \equiv \xi (z=z_-) 
=-1$ by assumption).  The field equation remains unchanged,
\be \label{eq-no-g-t}
\xi^2_t \xi_{tt} = \xi (\xi^2 -1)
\ee   
because the translation $t = z-z_-$ is a symmetry. Now we insert the power
series expansion
\be
\xi = -1  + b_2 t^2  + b_3 t^3 + \ldots
\ee
into the above equation (\ref{eq-no-g-t})  and find
\be \label{eq-b-2-t}
8 b_2^3 -2b_2 =0
\ee
with the three solutions 
\be \label{sol-b-2-t}
b_2 =0,\frac12 ,-\frac12 
\ee
where $b_2 =0$ corresponds to the
vacuum solution and $b_2 =\frac12$ corresponds to the compacton solution.
$b_2 = - \frac12$ corresponds to 
a solution growing in absolute value for increasing $t$,
which has infinite energy and is of no interest in this context. 
Observe, however, that $b_2 = -\frac12$ together with $\xi (t=0)=1$ would
correspond to the anti-compacton.
We remark that this degeneracy for the coefficient $b_2$ exists due to the
nonlinearity in the kinetic term and would be absent in a theory with normal
kinetic term. 
Once a choice
for $b_2$ has been made, the higher coefficients $b_n$ are determined
uniquely by linear equations. For the choice $b_2 = \frac12$, the higher
coefficients just reproduce the power series of the function $\xi (t) = - \cos
t$.    

Next, we study the power series expansion about the compacton center $z_0
=0$. By assumption, $\xi (z=0)=0$, therefore the expansion is
\be
\xi = b_1 z + b_2 z^2 + \ldots
\ee
If we assume that $b_1 \not= 0$, then it automatically follows that all even
$b_n$ are zero, $b_{n=2m}=0$. The odd $b_n$ are determined by linear equations
depending on the arbitrary nonzero $b_1$. For the choice $b_1 =1$ we recover
the power series expansion of the function $\xi (z) =\sin z$. It is easy to
see that finite energy requires, in fact, $b_1 = \pm 1$. Indeed, a first
integration of the field equation gives
\be
\xi_z^4 = (\xi ^2 -1)^2 + k
\ee
where $k$ is an integration constant. Finite energy requires $k=0$ which
implies for $\xi (z_0)=0$ that $\xi_z(z_0)=\pm 1$. 

Finally, let us briefly recapitulate the necessary conditions for the
existence of the compacton. These conditions are that
\begin{itemize}
\item[1)] the field may be joined smoothly with smooth first derivative to its
  vacuum value. This holds because of the triple degeneracy of 
$b_2=0,\frac12 ,-\frac12$ at the compacton boundaries $z_\pm$.
\item[2)] the field, which starts at the value $\xi (z_-)=-1$ at $z_-$ and 
obeys
  the compacton boundary condition $b_2 =\frac12$, evolves such that it takes
  the value $\xi (z_0)=0$ at some point $z_0$.  It may be checked easily that
  this property follows from the structure of the field equation.    
\item[3)] the field resulting from the evolution of point 2) is odd about the
  point $z_0$. This means that the evolution of $\xi$ for $z>z_0$ will be the
  mirror image of the evolution for $z<z_0$ and, consequently, will join the
  other vacuum, $\xi =1$, at $z_+ =2 z_0 -z_-$. This is automatically
  fulfilled in the case without gravitation, because a solution with initial
  condition $\xi (z_0 )=0$, $\xi_z (z_0) \not= 0$ is always an odd function
  about $z_0$ for arbitrary values of $\xi_z (z_0)$.
\end{itemize}

We remark that points 1) and 2) in the above list continue to hold in the case
with gravitation. On the other hand, point 3) will not hold in the generic
case. Instead, it requires a fine-tuning between the gravitational coupling
$\kappa$ and the cosmological constant $\Lambda$.

\section{The case with gravity}
\subsection{Expansion about the compacton boundary}

We again introduce the variable $t=z-z_-$ where by assumption 
$\xi (z= z_-)=\xi (t=0)=-1$, $\xi_t (t=0)=0$, and $C(t=0)=-\sqrt{\bar\Lambda
/3}$. The Einstein equations are
\br  
\frac34 C_{t} - C^2 &=& -\frac13 \bar \Lambda  + \kappa^2 (\xi^2 -1)^2 \\
\frac34 C_{t} &=& \kappa^2 \xi_t^4 , \label{eq-2-g-t}
\er
and we insert the power series expansion
\br \label{T-xi}
\xi &=& -1+b_2 t^2 + b_3 t^3 + b_4 t^4 + \ldots \\
\label{T-A}
C &=&  - \sqrt{\bar \Lambda /3} \,  + 
c_1 t + c_2 t^2 + \ldots  
\er
For $b_2$ we find again Equation (\ref{eq-b-2-t}) with the three solutions
(\ref{sol-b-2-t}), corresponding to vacuum, compacton and an infinite energy
solution,
respectively, like in the case without gravity. Choosing $b_2 = \frac12$ for
the compacton, the higher coefficients are, again, determined uniquely by
linear equations. Concretely we find 
\be
b_3 = -\frac{1}{15} \sqrt{\frac{\bar \Lambda}{3}}  \, , \quad
b_5 = -\sqrt{\frac{\bar \Lambda}{3}}\frac{1}{420}\left( \frac{16}{75}
\frac{\bar \Lambda}{3} +\frac{1}{2} \right)
\ee
\be
b_4 = - \frac{1}{24} + \frac{1}{90}\frac{\bar \Lambda}{3}  \, , \quad
b_6 =   \frac{223}{2\cdot 3^4
    \cdot 5^3 \cdot 7} \left(\frac{\bar \Lambda}{3}\right)^2
+ \frac{61}{2^4 \cdot 3^3 \cdot 5^2 \cdot 7} \frac{\bar \Lambda}{3}
+ \frac{1}{720}  .
\ee
These coefficients now depend on the cosmological constant and go to their
nongravitational values (the expansion coefficents of $-\cos t$) in the limit
$\bar \Lambda \to 0$, as they
should. 
As $b_3$ is negative, the cosmological constant tends to increase the size of
the compacton.
   
For the function $C$ we find from  Eq. (\ref{eq-2-g-t}) that the coefficients
$c_1$ - $c_4$ are identically zero. The two first nontrivial coefficients are
\be
c_5 = \frac{4\kappa^2}{15} \, ,\quad c_6 = \frac{8}{45}\kappa^2
\sqrt{\bar \Lambda /3} .
\ee

We remark that the first dependence of the $b_i$ on the  
 gravitational constant
$\kappa^2$ enters for the
coefficient $b_8$. 
The effect of the gravitational backreaction on the
power series expansion at the compacton boundary is, therefore, 
negligible for not too large values of the
gravitational constant. The contribution of $\kappa^2$ to $b_8$ is positive,
therefore the gravitational backreaction by itself 
tends to shrink the ``compacton'' radius. 

\subsection{Approximate determination of the compacton radius}

The fact that the expansion coefficients almost do not depend on the
gravitational constant $\kappa^2$ allows to determine the ``compacton'' radius
$|z_0 -z_-|$ for a given cosmological constant $\Lambda$ approximately. 
The idea is to use the power series expansion for
$\xi (t)$ to a certain order, and to determine the first zero of this
polynomial for $t>0$. We warn that for a fixed $\Lambda$ and arbitrary
$\kappa$ a true compacton does not exist. 
But the position of the first zero will
almost not depend on $\kappa$, therefore this position should determine the
compacton radius to a good approximation whenever the compacton exists, 
at least for not
too large values of $\Lambda$ and $\kappa$. Concretely, we will use the power
series expansion up to sixth order, where $\kappa$ does not show up at all.   

 Let us first check this proposal for the nongravitational
case, where the exact compacton solution is known. This nongravitational
solution is given in Eq. (\ref{compacton sol})
therefore the compacton radius $\rho = |z_0- z_-|$, variable $t$ and function 
$\xi (t)$ are
\be
\rho =\frac{\pi}{2} \, ,\quad t=\frac{\pi}{2} +z  \, ,\quad \xi (t) = -\cos t
\ee
Observe that all even derivatives of $\xi (t)$ at the compacton center
$t=\rho$ are zero, because $\xi $ must be an odd function of $z=t-\rho $.
Now we use the power series expansion of the function $\xi (t)= -\cos t$
about $t=0$ up to sixth order to determine the zero $\rho$ approximately, see
Eq. (\ref{table-1}). 
\be \label{table-1}
\begin{tabular}{r|c|c}

 $\xi (t)$ & compacton radius $\rho $ & $\xi'' (t=z_0)$  \\
\hline
 exact  ($-\cos t$) & $\pi /2 $ = 1.5708 & 0 \\
power series (${\cal O}(t^6)$) & 1.5699 & -0.0208 \\

\end{tabular} 
\ee
We find that the power series up to sixth order reproduces the exact value of
the compacton radius to a precision of better than $10^{-3}$. 
Further, we plot the value of the second
derivative at the compacton center,
$\xi '' (t=\rho )$, which must be zero for the exact solution.
This value is off about 2\% for the approximation. 

Now we do the same calculation for the power series up to sixth order using
the coefficients $b_i$ of the previous subsection for the case with gravitation
and cosmological constant included, for different values of the cosmological
constant. The result is presented in  Table 1.
It is clearly seen that the compacton radius increases with the cosmological
constant. 
Later we shall compare these simple results with a numerical integration and
will find that, indeed, they reproduce the compacton radius with a rather good
precision. We repeat that the existence of the compacton requires a fine-tuning
of the value of $\kappa$, but that the value of the ``compacton radius''
calculated above is completely insensitive to the value of $\kappa$.

\begin{center}
\begin{table}
\begin{tabular}{c|c}
cosmological constant $\bar \Lambda $ & compacton radius $\rho$    \\
\hline
0 & 1.5699   \\
0.0001 & 1.57145  \\
0.001 & 1.57481  \\
0.0025 & 1.57767  \\
0.005 & 1.58092  \\
0.01 & 1.58554  \\
0.025 & 1.59488  \\
0.05 & 1.6058  \\
0.1 & 1.62233 
\end{tabular} 
\caption{Approximate determination of the compacton radius from the power
  series expansion }
\end{table}
\end{center}

\subsection{Expansion about the center}

Now we want to study a power series expansion about $z_0$ where by definition
$\xi (z=z_0)=0$. Here we use the Einstein equations in the slightly rewritten
form
\br \label{ev-eq-1}
C_z &=& \frac43 \left( C^2 + \kappa^2 (\xi^2 -1)^2 -\frac{\bar \Lambda}{3}
\right) \\ \label{ev-eq-2}
\xi_z &=& \frac{1}{\sqrt{\kappa}} 
\left( C^2 + \kappa^2 (\xi^2 -1)^2 -\frac{\bar \Lambda}{3}
\right)^\frac14  
\er
where we choose the positive root for $\xi_z$, 
which describes the compacton (the
negative root would give the anticompacton). 
As said, for a compacton $\xi (z)$ must be an odd function about $z_0$. It may
be checked easily that a necessary and sufficient condition for this is that
$C$ is equal to zero at $z_0$, too, $C(z_0)=0$. Therefore, the conditions
at the center of a compacton are
\be
\xi(z_0) =0 \quad , \quad C(z_0)=0.
\ee
These equations already demonstrate the possible troubles for the existence of
a compacton. For suppose we start the evolution at the lower compacton
boundary $\xi (z_-)=-1$, $C(z_-)=-\sqrt{\bar \Lambda /3}$ and for compacton
boundary conditions (i.e. $b_2 = 1/2$). After some evolution $\xi (z) $ will
hit a point $z_0$ such that $\xi (z_0)=0$, but there is no reason that 
$C(z)$ has evolved to the same value $C(z_0)=0$ at $z=z_0$. The condition
$C(z_0)=0$ is an additional condition for the already completely determined
initial value problem (evolution from the compacton boundary with compacton
boundary conditions) and requires, therefore, a fine-tuning of the parameters
$\kappa$ and $\Lambda$. 

A first conclusion which may be drawn easily is that $\bar \Lambda >0$ for
$\kappa \not= 0$ is a necessary condition for the existence of a compacton.
For assume that we start the evolution of the fields at the center with
conditions   $\xi(z_0) =0$, $ C(z_0)=0$ for $\bar \Lambda =0$. The evolution
equations are
\br 
C_z &=& \frac43 \left( C^2 + \kappa^2 (\xi^2 -1)^2 
\right) \\
\xi_z &=& \frac{1}{\sqrt{\kappa}} 
\left( C^2 + \kappa^2 (\xi^2 -1)^2 
\right)^\frac14 . 
\er
From these equations and the conditions $C(z_0)=0$, $C_z (z_0)>0$ it follows
that $C(z)$ is strictly positive for $z>z_0$, and from this it follows in turn
that both $C_z$ and $\xi_z$ are strictly positive for all $z>z_0$. Therefore,
$\xi$ can never reach a point $z_+$ where $\xi_z (z_+)=0$, which is the
compacton boundary condition at the upper boundary of the compacton. 
The conclusion is that a compacton cannot exist for $\kappa \not= 0$ and
$\bar \Lambda =0$. 

An upper bound can be easily found for $\bar\Lambda$ by inserting the power
series expansion
\br
\xi &=& b_1 z + b_3 z^3 + \ldots \\
C &=& c_1 z + c_3 z^3 + \ldots
\er
into the above evolution equations (we assume momentarily that $z_0=0$).
It immediately follows that
\be
b_1 = \frac{1}{\sqrt{\kappa}}\left( \kappa^2 -\frac{\bar
    \Lambda}{3}\right)^\frac14
\ee
which is real only provided that
\be
\bar \Lambda \le 3\kappa^2 .
\ee     
Some further qualitative conclusions may be drawn by investigating the
evolution equations (\ref{ev-eq-1}), (\ref{ev-eq-2}), 
by assuming an evolution from the center $z_0$. The
problem is that $\xi_z$ must reach its value zero exactly at the same point
$z_+$ where $\xi$ takes the value $+1$. When $\xi_z$ has not yet reached zero
at the point $z_+$ where $\xi (z_+)=+1$, then it will never reach zero,
because from this point onward 
both quantities $C^2$ and $(\xi^2 -1)^2$ start to grow.  This happens for too
small $\bar \Lambda$ for a given $\kappa$.  
On the other hand, if $\xi_z$ takes its value zero before $\xi$ reaches the
value one, then all the higher derivatives of $\xi$ at this point are
singular, and the evolution equations break down at this point.    
This happens when $\bar \Lambda $ is too big for a given $\kappa$.
Therefore, for each given $\kappa \not= 0$, there exists precisely one fine
tuned value of $\bar \Lambda$ such that $\xi_z$ is zero exactly at the
point $z_+$ where $\xi$ takes the value $+1$, and, consequently, the compacton
exists only for this fine-tuned value. 

{\em Remark:} One may derive a single first order evolution equations for the
``orbit'' $C(\xi)$, and this equation may be used to determine the correlation
between $\bar \Lambda$ and $\kappa$ (but not for the compacton radius).
Indeed, dividing (\ref{ev-eq-1}) by (\ref{ev-eq-2}) one easily derives the
following equation,
\be \label{orbit-eq}
\tilde C_\xi = \beta \left[ \tilde C^2 -1 + \alpha (\xi^2
-1)^2 \right]^\frac{3}{4}
\ee
where
\be
\tilde C \equiv \sqrt{\frac{3}{\bar \Lambda}} C \, ,\quad
\alpha \equiv \frac{3 \kappa^2}{\bar \Lambda} \, , \quad
\beta \equiv \frac{4}{3} \left( \frac{\kappa^2 \bar \Lambda}{3}
\right)^\frac{1}{4} .
\ee
Between the compacton center and the upper compacton boundary (or, indeed,
between the lower and upper compacton boundary) both $C_z$ and $\xi_z$ are
strictly monotonous, therefore this ``orbit'' equation indeed may describe the
compacton. Integrating, e.g., from the center of the compacton to the upper
boundary requires the ``initial'' condition (at the center)
$\tilde C (\xi =0)=0 $ and the condition $\tilde C(\xi =1) =1$. These two
conditions for a first order ODE overdetermine the system and do not have a
solution in general.  Closer inspection of the above equation (\ref{orbit-eq})
again reveals that there exists exactly one $\beta$ for given $\alpha$ (or one
$\alpha$ for given $\beta$) such that the compacton exists. Indeed, as
$\tilde C_\xi$ always grows with growing $\alpha$ for fixed $\beta$
(and always grows with growing $\beta$ for fixed $\alpha$), independently of
the values of $\tilde C$ and $\xi$, the above statement follows. The numerical
values of $\alpha$ and $ \beta$ (and, therefore, of $\bar \Lambda$ and
$\kappa$) such that the compacton exists may be easily
determined numerically from Eq. (\ref{orbit-eq}) and completely agree with the
values determined from the numerical solution of the system of equations
 (\ref{ev-eq-1}) and (\ref{ev-eq-2}) presented in Table 2 and Fig. 1.  
We present the result of a numerical integration for some fine-tuned values
of $\kappa$ and $\bar \Lambda$ in the last figure (Fig. 15). Indeed, for these
fine-tuned values the function $\tilde C (\xi)$ reaches the value one at
$\xi =1$ to a high precision. 

\section{Numerical calculations}

\begin{center}
\begin{table}
\begin{tabular}{r|c|c|c}
 $\bar \Lambda $ & $\kappa^2$ &  $\rho_c$ & $\rho_b$   \\
\hline
$10^{-8}$ & 0.000073512465 & 1.56782 & 1.57096 \\
$10^{-6}$ & 0.00073531885 & 1.56817 & 1.57105 \\
$10^{-5}$ & 0.00232675114 & 1.57047 & 1.57127 \\
$10^{-4}$ & 0.00737252306 & 1.57080 & 1.57195 \\
0.001 & 0.0234612031 & 1.57251 & 1.57412 \\
0.005 & 0.0530591779 & 1.57515 & 1.57804 \\
0.01 & 0.0756746147 & 1.57896 & 1.58097 \\
0.025 & 0.1216666085 & 1.58443 & 1.59200 \\
0.05 & 0.1753053396 & 1.58917 & 1.59852 \\
0.1 & 0.2544852246 & 1.59367 & 1.60259 \\
0.2 & 0.3732506647 & 1.60912 & 1.61565 \\
0.3 & 0.4698973846 & 1.62347 & 1.62565 \\
0.5 & 0.6332261243 & 1.63673 & 1.64145 \\
0.7 & 0.7752995199 & 1.65032 & 1.65425 \\
1.0 & 0.9666302322 & 1.66660 & 1.67030 \\
1.5 & 1.252723578 & 1.68682 & 1.69224 \\
2.0 & 1.514998322 & 1.70648 & 1.71058 \\
3.0 & 1.999550771 & 1.73613 & 1.74101 \\
5.0 & 2.886273449 & 1.78229 & 1.78828
\end{tabular} 
\caption{Numerical determination of the fine-tuned value for $\kappa^2$ and of
the corresponding compacton radius. The displayed value of $\kappa^2$  for a 
given $\bar \Lambda$ is the lowest value of $\kappa^2$ that can be found 
numerically by a shooting from the center, 
such that a singularity does not form in the numerical
integration. Therefore, the true value of $\kappa^2$ could be slightly smaller,
and the value in the table is, in fact, an upper bound.
Correspondingly, the value of the compacton radius $\rho_c$, which is
also determined by a shooting from the center, is a lower bound. 
  The values for $\kappa^2$ are shown
to a very high precision (with many digits), because the determination of 
$\rho_c$  depends very sensitively on a precise
determination of $\kappa$. Finally, we display the compacton radius $\rho_b$
which is determined by a shooting from the boundary and depends on $\kappa^2$
only very weakly. }
\end{table}
\end{center}

Here we present the numerical calculations for the Einstein equations
(\ref{eq-1-g-z}),  (\ref{eq-2-g-z}). First we display the results for the
fine-tuned values of $\kappa$ for a given $\bar \Lambda$, as well as the
corresponding compacton radius, see Figures 1, 2 and Table\,2.  The fine-tuned
value of $\kappa$ for a given $\bar \Lambda$ is defined by the condition that
the numerical integration reproduces the compacton configuration. When
$\kappa$ is too big (or, equivalently, $\bar \Lambda$ is too small), then
$\xi_z$ will never reach the value zero. When
$\kappa$ is too small (or, equivalently, $\bar \Lambda$ is too big), then
$\xi_z$ reaches zero before $\xi$ reaches one. At this point a singularity
forms and the numerical integration breaks down. The fine-tuned value for
$\kappa$ is just the boundary case between these two cases.
More precisely, the values for $\kappa^2$ in Figure 1 and Table 2 are 
determined from a numerical integration which starts at the center of the
``compacton'' , that is, with initial values $C=0$, $\xi =0$ (``shooting from
the center''). Within this 
numerical intergration, the displayed values for $\kappa^2$ are the
smallest values for $\kappa^2$ that can be found numerically such that a
singularity does not form in the numerical integration.  
Strictly speaking, they are, therefore, upper bounds for the true fine-tuned
values of $\kappa^2$, but they are determined to a high precision.

In Table 2, we provide two values for the compacton radius. The first,
$\rho_c$, is determined by a shooting from the center, like $\kappa^2$
itself. 
The advantage of this value is that it provides a strict lower bound on the
compacton radius, because the ``compacton'' radius strictly decreases with
increasing $\kappa^2$ for a shooting from the center (here the ``compacton''
radius is defined as the point $z_+$
where $\xi (z_+)=1$, which is at the same time
the point where $\xi_z$ has a local minimum), and the used value for
$\kappa^2$ is an upper bound. The disadvantage of $\rho_c$ is that its value
is extremely sensitive to a precise determination of $\kappa^2$. An error in
the seventh of eighth digit for $\kappa^2$, for instance, frequently 
translates into an error in the third digit for $\rho_c$.    
Therefore, we determine the compacton radius in Table 2 also from a 
numerical integration which starts at the lower boundary of the compacton with
compacton boundary conditions (``shooting from the lower boundary'').
We know already from the power series analysis of the last section that the
so determined ``compacton'' radius $\rho_b$ almost does not depend on the exact
value of $\kappa^2$ (here the ``compacton'' radius is defined as the point
$t=z_0$ where $\xi(t=z_0)=0$). 
Therefore, we expect the so determined value $\rho_b$ to
better reproduce the true compacton radius than $\rho_c$. 
We find, for instance, that $\rho_b$ obeys $\rho_b > (\pi /2)$ for all $\bar
\Lambda$, which we know must be true at least for small $\bar \Lambda$ (see
the power series expansion analysis of the last section). 
The disadvantage of
$\rho_b$ is that it no longer provides a strict lower bound, therefore we
display both values.   
  
Next, we present figures with some details from the numerical integration.
We do the numerical integration in two different ways. First, we start at the
center where both $\xi$ and $C$ are zero (``shooting from the center''), see
Figures 3-8. We fix $\bar \Lambda =0.5$ and choose for $\kappa$ the fine-tuned
value as well as one slightly smaller and one slightly bigger value. The
results for other values of $\bar \Lambda$ are completely equivalent, and we
choose the value of $0.5$ for $\bar \Lambda$ just for convenience.

Then, we present the corresponding figures which one obtains for the
numerical integration which starts at the lower boundary of the compacton
(more precisely, very near the lower boundary, where the compacton boundary
condition $b_2=1/2$ is taken into account), and integrates all the way up to
the upper boundary (``shooting from the lower boundary''), see Figures 9-14. 
We remark that from these figures one may see explicitly that the ``compacton
radius'' $\rho_b$ determined from  the shooting from the lower boundary
depends only weakly on $\kappa^2$, and therefore, approximates the true
compacton radius rather well even if $\kappa^2$ is not exactly equal to its
fine-tuned value. Indeed, from Figures 9, 11, and 13 we find the following
values for $\rho_b$, given $\bar \Lambda =0.5$ and the values for $\kappa^2$
indicated in the table,
\be \label{table-4}
\begin{tabular}{r|c|c}
$ \bar \Lambda $ & $\kappa^2$   & $\rho_b $  \\
\hline
 0.5   & 0.633226   &   1.64145 \\
0.5 & 0.608400  & 1.64236 \\
0.5 & 0.640000 & 1.64121
\end{tabular} 
\ee
where the first value $\kappa^2 = 0.633226$ is the fine-tuned value. We find
that when $\kappa^2$ has a relative deviation of about $10^{-2}$ from its 
fine-tuned value, then the relative deviation of $\rho_b$ is about 
$2\cdot 10^{-5}$, i.e., quite small, indeed. 

Finally, we present the result of a numerical integration of the ``orbit''
equation (\ref{orbit-eq}) for some fine-tuned values of $\kappa$ and 
$\bar \Lambda$, with initial condition $\tilde C(\xi =0)=0$. For the
fine-tuned values $\tilde C$ reaches the value
$\tilde C(\xi=1)=1$ to a high precision. 

We conclude that the 
numerical analysis completely confirms the qualitative discussion 
of the last section, both for an evolution from the center and an evolution
from the boundary of the compacton.

\section{Linear Stability}

In this section we want to discuss the issue of stability of the compacton
domain wall under linear fluctuations. We shall not further comment on
the issue of stability under gravitational fluctuations. Stability under
fluctuations of the gravitational field for a whole class of thick brane
models has been extensively discussed in Ref. \cite{CEHS1}, and we
demonstrated already in Ref. \cite{AGSGW1} that our compact domain wall
solutions belong to that class and, therefore, the analysis of Ref. \cite{CEHS1}
applies. 

On the other hand, we want to study linear stability under fluctuations of the
scalar field for the system with gravitational backreaction (for the compacton
without gravity, linear stability has already beed demonstrated in Ref.
\cite{AGSGW1}). It turns out that stability of the full system with gravity 
may indeed be demonstrated by a combination of analytical and numerical
methods, at least for not too large values of $\bar \Lambda$ (and,
consequently, also for not too large values of $\kappa$). The restriction to
not too large values of $\bar \Lambda$ probably does not mean that stability
does not hold for large values, but just indicates that an approximation
used in our proof becomes too crude for large values of $\bar \Lambda$.

The linear fluctuation equation for the scalar field is
\be
\xi_z^2 \eta_{zz} +2 (\xi_z \xi_{zz} -A_z \xi_z^2 )\eta_z -(3\xi^2 -1)\eta =
\frac13 e^A \xi_z^2 \Box \eta
\ee
where $\Box$ is the four-dimensional Minkowski space wave operator, 
and $A$ and $\xi$ are the
compact brane solutions of Sections 4, 5. 
Further, $\eta$ is the linear fluctuation.
Outside the compacton region, where the
fields $\xi$ and $A$ take their vacuum values (specifically, $\xi_z=0$),
the only possible solution is $\eta =0$, like in the non-gravitational case.
For the case inside the compacton region, we use
the separation of 
variable ansatz $\eta \to \eta (z) \phi (x)$ and get
\be
\Box \phi =-\omega^2 \phi
\ee
(the Klein--Gordon equation) and
\be \label{eta-eq1}
\eta_{zz} +2(\frac{\xi_{zz}}{\xi_z} - A_z )\eta_z -\frac{3\xi^2 -1}{\xi_z^2} 
\eta = -\omega^2 e^A \eta
\ee
where $\omega^2$ is the separation constant.
After the transformation $\eta = \frac{e^A}{\xi_z} \bar \eta $,
Equation (\ref{eta-eq1}) can be brought into the Schroedinger equation form
\be
-\bar \eta_{zz} + U(z) \bar \eta =
\omega^2 e^A \bar \eta
\ee
where 
\be
U(z)\equiv  
A_z^2 - A_{zz} + \frac{\xi_{zzz}}{\xi_z} -
2 A_z \frac{\xi_{zz}}{\xi_z} +\frac{3\xi^2 -1}{\xi_z^2} .
\ee
We now eliminate all higher derivatives terms (first and higher derivatives 
of $\xi$, second derivative of $A$) with the help of the field equations for 
$\xi$ and $A$. Further, we again use the notation $C(z) \equiv A_z$,
$C=\frac{\bar \Lambda}{3} \bar C$ and find
\bdi
U=-\frac{4}{9} \Gamma + \frac{\bar \Lambda}{27}\bar C^2 + \frac{8}{3}
\kappa^\frac{3}{2} \sqrt{\frac{\bar \Lambda}{3}} \bar C \xi 
\frac{1-\xi^2}{\sqrt{\Gamma}} \Gamma^{-\frac14 } -
\edi
\be
- 2\kappa \frac{1-\xi^2}{\sqrt{\Gamma}} +4\kappa \xi^2 \Gamma^{-\frac12 }
-2\kappa^3 \frac{(1-\xi^2)^2}{\Gamma}\xi^2 \Gamma^{-\frac12 }
\ee
where 
\be
\Gamma = \frac{\bar \Lambda }{3} (\bar C^2 -1) + \kappa^2 (\xi^2 -1)^2 .
\ee
Stability requires $\omega^2 \ge 0$. This, in turn, follows from the positive 
semi-definiteness of the above Schroedinger operator
\be
H\equiv -\partial_z^2 +U
\ee
because from $\langle \eta |H|\eta \rangle \ge 0$ and the obvious
 $\langle \eta |e^A|\eta \rangle > 0$ it follows that $\omega^2 \ge 0$.  

Next, let us study some properties of the potential $U$. $U$ tends to
plus infinity at the compacton boundaries $z=\pm \rho $ (here we assume that 
the compacton center is at $z=0$). This follows from the fact that 
$\Gamma$ goes to zero
at the compacton boundaries, and from the fact that
$$ 
\lim_{z\to \pm \rho} \frac{(1-\xi^2)^2}{\Gamma} = \kappa^{-2}
$$
Further, $U$ takes its only minimum at the center $z=0$, where $\xi(z=0)=0$
and $C(z=0)=0$. The value at
the minimum is
\be \label{U_0}
U_0 \equiv U(z=0) = - \frac{4}{9} (\kappa^2 - \frac{\bar \Lambda}{3})
- \frac{2\kappa}{ (\kappa^2 - \frac{\bar \Lambda}{3})^\frac{1}{2}}
\ee
The fact that $U$ has an extremum at the center follows from the symmetry of
$U$ under the reflection $z \to -z$, whereas the fact that this is the only
extremum (minimum) of $U$ is not so easy to prove, because both the
compacton solutions $\xi $ and $C$ and the correlation between $\bar \Lambda$
and $\kappa$ are needed to demonstrate it. Here we restrict ourselves to a
numerical demonstration. The potential $U$ is plotted for the three values
$\bar \Lambda =  10^{-3}, 0.5, 6.0$ in Figure 16. The behaviour described
above is clearly seen in these figures. It follows that the Schroedinger
operator with potential $U$ has a discrete, non-degenerate 
spectrum of eigenvalues
\be
H\psi_n = E_n \psi_n \quad ,\qquad E_1 < E_2 < E_3 < \ldots
\ee
and we want to demonstrate that $E_1 =0$, $E_n >0$ for $n=2,3,\ldots$.
Observe that we have to expect one zero
eigenvalue because of the Goldstone mode of the translational symmetry
$z \to z+a$.  

The idea now is to find a new potential $\bar U$ which is less or equal to
the potential $U$ for all $z$ in the compacton region,
\be
\bar U(z) \le U(z) \quad , \qquad z \in [-\rho ,\rho ]
\ee
and to study the spectral problem
\be
\bar H \bar \psi_n \equiv
(-\partial_z^2 + \bar U)\bar \psi_n = \bar E_n \bar \psi_n
\quad ,\qquad \bar E_1 < \bar E_2 < \bar E_3 < \ldots
\ee
for this new potential $\bar U$. Stability is proved if the new potential
has precisely one negative eigenvalue $\bar E_1$, whereas all the other
eigenvalues $\bar E_n, n\ge 2$ are positive. This follows from the inequality
(which we prove below, see Eq. (\ref{ineq-proof}))
\be \label{energy-ineq}
E_n \ge \bar E_n \quad \forall \quad n
\ee
because then $E_2 \ge \bar E_2 >0$ and, consequently, only the lowest
eigenvalue $E_1$ can be less or equal to zero. Finally, we know that it must be
equal to zero, $E_1 =0$, because we know that a zero egenvalue must exist
(the above-mentioned Goldstone mode of the translational symmetry
$z \to z+a$).  Therefore, the Schroedinger operator $H$ cannot have a negative 
eigenvalue, which is what we wanted to prove.

Concretely, we choose the infinite square well potential for $\bar U$, 
which is plus
infinity at the compacton boundaries, and equal to $U_0$ in the interior.  
The spectrum for this potential is well-known and is given by the
eigenvalues
\be
\bar E_n = \frac{\pi^2 \hbar^2}{2mL^2} n^2 +U_0
\ee
where $L$ is the width of the square well.
In our case $\frac{\hbar^2}{2m}=1$ and $L=2\rho$ and, therefore,
\be
\bar E_n = \frac{\pi^2}{4\rho^2} n^2 +U_0 .
\ee 
Stability requires $\bar E_1 <0$ and $\bar E_2 >0$ or
\br
\bar E_1 \equiv \frac{\pi^2}{4\rho^2} + U_0 &<& 0 \\  
\bar E_2 \equiv \frac{\pi^2}{\rho^2} + U_0 &>& 0
\er
where $U_0$ is given in Eq. (\ref{U_0}) and for $\rho$, $\bar \Lambda$ and
$\kappa$ the values of Table 2 have to be used. 

Qualitatively we can observe at once from Table 2 that our stability proof 
cannot work for arbitrarily large $\bar \Lambda$, because $|U_0|$ grows
with growing $\bar \Lambda$, whereas $\rho^{-2}$ shrinks with growing
$\bar \Lambda $. The numerical results are shown in Figure 17 (for $\bar E_1$)
and in Figure 18 (for $\bar E_2$). We find that $\bar E_1 <0$ for all 
$\bar \Lambda $
as must hold because of the existence of the Goldstone mode.
For $\bar E_2$ we find that $\bar E_2 >0$ at least for 
$0 \le \bar \Lambda \le 2$,
therefore we have proved stability for this parameter region.

{\em Remark:} we could, in principle, determine the value of $\bar \Lambda$ 
where the transition from positive to negative $\bar E_2$ occurs with higher
precision (it will happen very near to $\bar \Lambda =3$). But this would
require the numerical solution of the compacton-gravity system for these
values of $\bar \Lambda$, which is quite demanding and does not produce 
important new insights, therefore we do not perform this calculation here.

{\em Remark:} the fact that our stability proof does not work for high values
of $\bar \Lambda$ does not necessarily mean that stability does not hold for
these high values. It could just be the case that the approximation of the true
potential $U$ by the infinite square well $\bar U$ is worse for higher values
of $\bar \Lambda$. This is indeed indicated by the potentials $U$ plotted in
Figure 16. The higher the value of $\bar \Lambda$, the steeper the valley
in the center of the potential, and the  worse the approximation by a flat
line.   
 
{\em Remark:} the stability proof also works in the limit $\bar \Lambda \to 0$,
$\kappa \to 0$, that is, in the non-gravitational case. This may be seen
most directly by using the non-gravitational compacton
solution $\rho = \frac{\pi}{2}$,
$\xi =\sin z$ and the resulting potential
$U = -2 + 2\tan^2 z$, which obviously may be bound by an infinite square wall
potential with depth $U_0 = -2$. Therefore, our stability proof also provides
an alternative proof for the stability of the non-gravitational
compacton of Ref. \cite{ASGW1}, \cite{AGSGW1}. 

Finally, we still have to prove the inequality (\ref{energy-ineq}) which we
used in our stability proof. The proof only uses elementary quantum mechanics,
but as we could not find it in any of the standard text books, we provide it
here for completeness.

{\em Theorem:}
Given the two Schroedinger operators $H=-\partial_z^2 +U(z)$ and $\bar H=
-\partial_z^2 + \bar U(z)$ where $U(z)\ge U_0 \; \forall \; z 
\in (-\rho ,\rho )$,
$\quad \bar U(z) \ge \bar U_0 \; \forall \; z \in (-\rho ,\rho )$, 
$\quad \lim_{z \to \pm \rho} U(z) = +\infty$ and   $\lim_{z \to \pm \rho}
\bar U(z) = +\infty$ such that the spectrum of both $H$ and $\bar H$ is
purely discrete and non-degenerate,
$$
H\psi_n = E_n \psi_n \quad ,\qquad E_1 < E_2 < E_3 < \ldots
$$
$$
\bar H \bar \psi_n = \bar E_n \bar \psi_n \quad ,\qquad 
\bar E_1 < \bar E_2 < \bar E_3 < \ldots
$$
and further it holds that 
$$
U(z) \ge \bar U(z) \quad \forall \quad z \in ( -\rho ,\rho )
$$
then the inequality
$$
E_n \ge \bar E_n \quad \forall \quad n
$$
holds.

{\em Proof:}
We define the Schroedinger operator
\be \label{ineq-proof}
H(s) \equiv -\partial_z^2 + \bar U + s (U-\bar U)
\ee
which has a purely discrete and non-degenerate spectrum
\be
H(s) \psi_n (s) = E_n (s) \psi_n (s) 
\ee
for all $s \in [0,1]$. It follows that 
\be
E_n (s) = \langle \psi_n (s) | H(s) | \psi_n (s) \rangle
\ee
and, after a derivative w.r.t. $s$, and using that
$$
\langle  (d/ds) \psi_n (s) |  \psi_n (s) \rangle +
\langle \psi_n (s) | (d/ds) \psi_n (s) \rangle =0,
$$ 
we further get
\be 
\frac{d}{ds} E_n (s) = \langle \psi_n (s) | (U-\bar U) | 
\psi_n (s) \rangle \ge 0 ,
\ee
from which the theorem follows immediately. QED.

\section{Discussion}
In this paper, we completed the investigation which we started in 
\cite{AGSGW1}, studying in detail the existence of compact domain wall
solutions in the full system of a quartic {\em K} field coupled to gravity.
The result of this investigation is that such compact domain walls do exist,
but their existence requires a correlation between the 
gravitational coupling $\kappa$ and the bulk cosmological constant
$\Lambda$. Further, these compact domain walls are stable under linear
perturbations. 
Firstly,
we want to emphasize that the correlation between the
gravitational coupling $\kappa$ and the bulk cosmological constant
$\Lambda$ we find in our system 
is qualitatively different from a fine-tuning which is frequently found in
three-brane models. There, a fine-tuning between the bulk cosmological
constant and the brane tension (or some related parameters of the brane) is
necessary in order to achieve a vanishing effective four-dimensional
cosmological constant. In our case, on the other hand, the full action 
density is always zero on-shell locally
(i.e., for any local solution of the Einstein
equations). The correlation is needed, instead, to guarantee the global
existence of a solution with finite energy in the transverse direction
in the non-gravitational sector
alone, that is, the global existence of a topological defect or domain wall. 
This correlation is, in fact, a rather interesting observation by itself. One
may see it as a drawback, requiring a ``fine-tuning'' of the cosmological
constant for a given gravitational constant, or as an advantage, providing a
``prediction'' of the bulk cosmological constant from the value of the
gravitational
constant and the requirement of the existence of a thick brane universe.
Quantitatively, the diameter of of the brane in the transverse direction grows
with growing cosmological constant, and this growth may be seen as the net
result of two competing influences. The gravitational constant by itself tends
to shrink the brane size, whereas the cosmological constant tends to 
increase it. But these two constants are correlated, and the influence of the
cosmological constant is stronger. The net effect is, therefore, an increase
in the size of the brane. 

Another interesting point consists in the fact that we were able to go rather
far in the analysis of the model, establishing both the existence 
and the stability of the
compact domain wall solutions, as well as its numerical properties to a high
precision. Due to the inherent nonlinearity of the gravitational backreaction,
such systems are usually quite difficult to analyse, and some simplifying
assumptions (like replacing the ``matter'' fields by some effective energy
momentum distribution) are frequently employed. In the present paper, on the
other hand, we performed a full field theory calculation without any of these
simplifying assumptions.   

Further, let us emphasize that the results of this paper should be easily
generalizable to other models. In fact, all that is needed is the presence of
a potential term for the scalar {\em K} field which allows for a vacuum
degeneracy, and the presence of a generalized kinetic term which takes a
certain nonstandard form in the limit of low energy. Still, this latter
condition is rather nontrivial, because the absence of the standard quadratic
kinetic
term at low energies is a necessary condition. In this sense these models are
special, because the presence of the quadratic term is a typical feature of
low energy effective theories. Under which conditions these special theories
may be induced remains an open problem at the moment, which deserves further
study. In any case, the fact that essential features of brane physics, like
domain walls of compact support, suppression of linear propagation outside the
domain wall, and standard linear propagation inside the brane, are naturally
provided by these models, makes it worth studying these special theories.

Finally, we want to mention that recently {\em K} field theories have been
used within the context of brane physics for a slightly different purpose in
\cite{Ol1}. In that paper it was shown that the distance between different
(thin) branes may be stabilized if the bulk scalar field has a nonstandard
kinetic term.

\section*{Acknowledgements}

C.A., P.K. and J.S.-G. thank MCyT (Spain) and FEDER
(FPA2005-01963), and support from Xunta de Galicia (grant
PGIDIT06PXIB296182PR and Conselleria de Educacion). 
A.W. acknowledges support from the Foundation for Polish
Science FNP (KOLUMB programme) and Ministry of Science and Higher
Education of Poland (grant N N202 126735).
N.E.G. thanks  Xunta de Galicia for support and Departamento
de Fisica de
Particulas - Universidade de Santiago de Compostela for hospitality
during this work.
Further, the authors thank O. Alvarez for helpful correspondence related to
the stability proof of Section 6.

\newpage

\def\temp{1.34}%
\let\tempp=\relax
\expandafter\ifx\csname psboxversion\endcsname\relax
  \message{PSBOX(\temp) loading}%
\else
    \ifdim\temp cm>\psboxversion cm
      \message{PSBOX(\temp) loading}%
    \else
      \message{PSBOX(\psboxversion) is already loaded: I won't load
        PSBOX(\temp)!}%
      \let\temp=\psboxversion
      \let\tempp= 
    \fi
\fi
\tempp
\let\psboxversion=\temp
\catcode`\@=11
%
%
\def\psfortextures{
\def\PSspeci@l##1##2{%
\special{illustration ##1\space scaled ##2}%
}}%
\def\psfordvitops{
\def\PSspeci@l##1##2{%
\special{dvitops: import ##1\space \the\drawingwd \the\drawinght}%
}}%
\def\psfordvips{
\def\PSspeci@l##1##2{%
\d@my=0.1bp \d@mx=\drawingwd \divide\d@mx by\d@my
\includegraphics{##1\space}}}%
\def\psforoztex{
\def\PSspeci@l##1##2{%
\special{##1 \space
      ##2 1000 div dup scale
      \number-\psllx\space \number-\pslly\space translate
}}}%
\def\psfordvitps{
\def\psdimt@n@sp##1{\d@mx=##1\relax\edef\psn@sp{\number\d@mx}}
\def\PSspeci@l##1##2{%
\special{dvitps: Include0 "psfig.psr"}
\psdimt@n@sp{\drawingwd}
\special{dvitps: Literal "\psn@sp\space"}
\psdimt@n@sp{\drawinght}
\special{dvitps: Literal "\psn@sp\space"}
\psdimt@n@sp{\psllx bp}
\special{dvitps: Literal "\psn@sp\space"}
\psdimt@n@sp{\pslly bp}
\special{dvitps: Literal "\psn@sp\space"}
\psdimt@n@sp{\psurx bp}
\special{dvitps: Literal "\psn@sp\space"}
\psdimt@n@sp{\psury bp}
\special{dvitps: Literal "\psn@sp\space startTexFig\space"}
\special{dvitps: Include1 "##1"}
\special{dvitps: Literal "endTexFig\space"}
}}%
\def\psfordvialw{
\def\PSspeci@l##1##2{
\special{language "PostScript",
position = "bottom left",
literal "  \psllx\space \pslly\space translate
  ##2 1000 div dup scale
  -\psllx\space -\pslly\space translate",
include "##1"}
}}%
\def\psforptips{
\def\PSspeci@l##1##2{{
\d@mx=\psurx bp
\advance \d@mx by -\psllx bp
\divide \d@mx by 1000\multiply\d@mx by \xscale
\incm{\d@mx}
\let\tmpx\dimincm
\d@my=\psury bp
\advance \d@my by -\pslly bp
\divide \d@my by 1000\multiply\d@my by \xscale
\incm{\d@my}
\let\tmpy\dimincm
\d@mx=-\psllx bp
\divide \d@mx by 1000\multiply\d@mx by \xscale
\d@my=-\pslly bp
\divide \d@my by 1000\multiply\d@my by \xscale
\at(\d@mx;\d@my){\special{ps:##1 x=\tmpx, y=\tmpy}}
}}}%
\def\psonlyboxes{
\def\PSspeci@l##1##2{%
\at(0cm;0cm){\boxit{\vbox to\drawinght
  {\vss\hbox to\drawingwd{\at(0cm;0cm){\hbox{({\tt##1})}}\hss}}}}
}}%
\def\psloc@lerr#1{%
\let\savedPSspeci@l=\PSspeci@l%
\def\PSspeci@l##1##2{%
\at(0cm;0cm){\boxit{\vbox to\drawinght
  {\vss\hbox to\drawingwd{\at(0cm;0cm){\hbox{({\tt##1}) #1}}\hss}}}}
\let\PSspeci@l=\savedPSspeci@l
}}%
%
%
\newread\pst@mpin
\newdimen\drawinght\newdimen\drawingwd
\newdimen\psxoffset\newdimen\psyoffset
\newbox\drawingBox
\newcount\xscale \newcount\yscale \newdimen\pscm\pscm=1cm
\newdimen\d@mx \newdimen\d@my
\newdimen\pswdincr \newdimen\pshtincr
\let\ps@nnotation=\relax
{\catcode`\|=0 |catcode`|\=12 |catcode`|
|catcode`#=12 |catcode`*=14
|xdef|backslashother{\}*
|xdef|percentother{
|xdef|tildeother{~}*
|xdef|sharpother{#}*
}%
\def\R@moveMeaningHeader#1:->{}%
\def\uncatcode#1{%
\edef#1{\expandafter\R@moveMeaningHeader\meaning#1}}%
\def\execute#1{#1}
\def\psm@keother#1{\catcode`#112\relax}
\def\executeinspecs#1{%
\execute{\begingroup\let\do\psm@keother\dospecials\catcode`\^^M=9#1\endgroup}}%
\def\@mpty{}%
\def\matchexpin#1#2{
  \fi%
  \edef\tmpb{{#2}}%
  \expandafter\makem@tchtmp\tmpb%
  \edef\tmpa{#1}\edef\tmpb{#2}%
  \expandafter\expandafter\expandafter\m@tchtmp\expandafter\tmpa\tmpb\endm@tch%
  \if\match%
}%
\def\matchin#1#2{%
  \fi%
  \makem@tchtmp{#2}%
  \m@tchtmp#1#2\endm@tch%
  \if\match%
}%
\def\makem@tchtmp#1{\def\m@tchtmp##1#1##2\endm@tch{%
  \def\tmpa{##1}\def\tmpb{##2}\let\m@tchtmp=\relax%
  \ifx\tmpb\@mpty\def\match{YN}%
  \else\def\match{YY}\fi%
}}%
\def\incm#1{{\psxoffset=1cm\d@my=#1
 \d@mx=\d@my
  \divide\d@mx by \psxoffset
  \xdef\dimincm{\number\d@mx.}
  \advance\d@my by -\number\d@mx cm
  \multiply\d@my by 100
 \d@mx=\d@my
  \divide\d@mx by \psxoffset
  \edef\dimincm{\dimincm\number\d@mx}
  \advance\d@my by -\number\d@mx cm
  \multiply\d@my by 100
 \d@mx=\d@my
  \divide\d@mx by \psxoffset
  \xdef\dimincm{\dimincm\number\d@mx}
}}%
%
\newif\ifNotB@undingBox
\newhelp\PShelp{Proceed: you'll have a 5cm square blank box instead of
your graphics (Jean Orloff).}%
\def\s@tsize#1 #2 #3 #4\@ndsize{
  \def\psllx{#1}\def\pslly{#2}%
  \def\psurx{#3}\def\psury{#4}
  \ifx\psurx\@mpty\NotB@undingBoxtrue
  \else
    \drawinght=#4bp\advance\drawinght by-#2bp
    \drawingwd=#3bp\advance\drawingwd by-#1bp
  \fi
  }%
\def\sc@nBBline#1:#2\@ndBBline{\edef\p@rameter{#1}\edef\v@lue{#2}}%
\def\g@bblefirstblank#1#2:{\ifx#1 \else#1\fi#2}%
{\catcode`\%=12
\xdef\B@undingBox{
\def\ReadPSize#1{
 \readfilename#1\relax
 \let\PSfilename=\lastreadfilename
 \openin\pst@mpin=#1\relax
 \ifeof\pst@mpin \errhelp=\PShelp
   \errmessage{I haven't found your postscript file (\PSfilename)}%
   \psloc@lerr{was not found}%
   \s@tsize 0 0 142 142\@ndsize
   \closein\pst@mpin
 \else
   \if\matchexpin{\GlobalInputList}{, \lastreadfilename}%
   \else\xdef\GlobalInputList{\GlobalInputList, \lastreadfilename}%
     \immediate\write\psbj@inaux{\lastreadfilename,}%
   \fi%
   \loop
     \executeinspecs{\catcode`\ =10\global\read\pst@mpin to\n@xtline}%
     \ifeof\pst@mpin
       \errhelp=\PShelp
       \errmessage{(\PSfilename) is not an Encapsulated PostScript File:
           I could not find any \B@undingBox: line.}%
       \edef\v@lue{0 0 142 142:}%
       \psloc@lerr{is not an EPSFile}%
       \NotB@undingBoxfalse
     \else
       \expandafter\sc@nBBline\n@xtline:\@ndBBline
       \ifx\p@rameter\B@undingBox\NotB@undingBoxfalse
         \edef\t@mp{%
           \expandafter\g@bblefirstblank\v@lue\space\space\space}%
         \expandafter\s@tsize\t@mp\@ndsize
       \else\NotB@undingBoxtrue
       \fi
     \fi
   \ifNotB@undingBox\repeat
   \closein\pst@mpin
 \fi
\message{#1}%
}%
%
%
\def\psboxto(#1;#2)#3{\vbox{%
   \ReadPSize{#3}%
   \advance\pswdincr by \drawingwd
   \advance\pshtincr by \drawinght
   \divide\pswdincr by 1000
   \divide\pshtincr by 1000
   \d@mx=#1
   \ifdim\d@mx=0pt\xscale=1000
         \else \xscale=\d@mx \divide \xscale by \pswdincr\fi
   \d@my=#2
   \ifdim\d@my=0pt\yscale=1000
         \else \yscale=\d@my \divide \yscale by \pshtincr\fi
   \ifnum\yscale=1000
         \else\ifnum\xscale=1000\xscale=\yscale
                    \else\ifnum\yscale<\xscale\xscale=\yscale\fi
              \fi
   \fi
   \divide\drawingwd by1000 \multiply\drawingwd by\xscale
   \divide\drawinght by1000 \multiply\drawinght by\xscale
   \divide\psxoffset by1000 \multiply\psxoffset by\xscale
   \divide\psyoffset by1000 \multiply\psyoffset by\xscale
   \global\divide\pscm by 1000
   \global\multiply\pscm by\xscale
   \multiply\pswdincr by\xscale \multiply\pshtincr by\xscale
   \ifdim\d@mx=0pt\d@mx=\pswdincr\fi
   \ifdim\d@my=0pt\d@my=\pshtincr\fi
   \message{scaled \the\xscale}%
 \hbox to\d@mx{\hss\vbox to\d@my{\vss
   \global\setbox\drawingBox=\hbox to 0pt{\kern\psxoffset\vbox to 0pt{%
      \kern-\psyoffset
      \PSspeci@l{\PSfilename}{\the\xscale}%
      \vss}\hss\ps@nnotation}%
   \global\wd\drawingBox=\the\pswdincr
   \global\ht\drawingBox=\the\pshtincr
   \global\drawingwd=\pswdincr
   \global\drawinght=\pshtincr
   \baselineskip=0pt
   \copy\drawingBox
 \vss}\hss}%
  \global\psxoffset=0pt
  \global\psyoffset=0pt
  \global\pswdincr=0pt
  \global\pshtincr=0pt 
  \global\pscm=1cm 
}}%
%
%
\def\psboxscaled#1#2{\vbox{%
  \ReadPSize{#2}%
  \xscale=#1
  \message{scaled \the\xscale}%
  \divide\pswdincr by 1000 \multiply\pswdincr by \xscale
  \divide\pshtincr by 1000 \multiply\pshtincr by \xscale
  \divide\psxoffset by1000 \multiply\psxoffset by\xscale
  \divide\psyoffset by1000 \multiply\psyoffset by\xscale
  \divide\drawingwd by1000 \multiply\drawingwd by\xscale
  \divide\drawinght by1000 \multiply\drawinght by\xscale
  \global\divide\pscm by 1000
  \global\multiply\pscm by\xscale
  \global\setbox\drawingBox=\hbox to 0pt{\kern\psxoffset\vbox to 0pt{%
     \kern-\psyoffset
     \PSspeci@l{\PSfilename}{\the\xscale}%
     \vss}\hss\ps@nnotation}%
  \advance\pswdincr by \drawingwd
  \advance\pshtincr by \drawinght
  \global\wd\drawingBox=\the\pswdincr
  \global\ht\drawingBox=\the\pshtincr
  \global\drawingwd=\pswdincr
  \global\drawinght=\pshtincr
  \baselineskip=0pt
  \copy\drawingBox
  \global\psxoffset=0pt
  \global\psyoffset=0pt
  \global\pswdincr=0pt
  \global\pshtincr=0pt 
  \global\pscm=1cm
}}%
%
\def\psbox#1{\psboxscaled{1000}{#1}}%
\newif\ifn@teof\n@teoftrue
\newif\ifc@ntrolline
\newif\ifmatch
\newread\j@insplitin
\newwrite\j@insplitout
\newwrite\psbj@inaux
\immediate\openout\psbj@inaux=psbjoin.aux
\immediate\write\psbj@inaux{\string\joinfiles}%
\immediate\write\psbj@inaux{\jobname,}%
%
%
\def\toother#1{\ifcat\relax#1\else\expandafter%
  \toother@ux\meaning#1\endtoother@ux\fi}%
\def\toother@ux#1 #2#3\endtoother@ux{\def\tmp{#3}%
  \ifx\tmp\@mpty\def\tmp{#2}\let\next=\relax%
  \else\def\next{\toother@ux#2#3\endtoother@ux}\fi%
\next}%
%
%
\let\readfilenamehook=\relax
\def\re@d{\expandafter\re@daux}
\def\re@daux{\futurelet\nextchar\stopre@dtest}%
\def\re@dnext{\xdef\lastreadfilename{\lastreadfilename\nextchar}%
  \afterassignment\re@d\let\nextchar}%
\def\stopre@d{\egroup\readfilenamehook}%
\def\stopre@dtest{%
  \ifcat\nextchar\relax\let\nextread\stopre@d
  \else
    \ifcat\nextchar\space\def\nextread{%
      \afterassignment\stopre@d\chardef\nextchar=`}%
    \else\let\nextread=\re@dnext
      \toother\nextchar
      \edef\nextchar{\tmp}%
    \fi
  \fi\nextread}%
\def\readfilename{\bgroup%
  \let\\=\backslashother \let\%=\percentother \let\~=\tildeother
  \let\#=\sharpother \xdef\lastreadfilename{}%
  \re@d}%
%
%
\xdef\GlobalInputList{\jobname}%
\def\psnewinput{%
  \def\readfilenamehook{
    \if\matchexpin{\GlobalInputList}{, \lastreadfilename}%
    \else\xdef\GlobalInputList{\GlobalInputList, \lastreadfilename}%
      \immediate\write\psbj@inaux{\lastreadfilename,}%
    \fi%
    \ps@ldinput\lastreadfilename\relax%
    \let\readfilenamehook=\relax%
  }\readfilename%
}%
\expandafter\ifx\csname @@input\endcsname\relax    
  \immediate\let\ps@ldinput=\input\def\input{\psnewinput}%
\else
  \immediate\let\ps@ldinput=\@@input
  \def\@@input{\psnewinput}%
\fi%
\def\nowarnopenout{%
 \def\warnopenout##1##2{%
   \readfilename##2\relax
   \message{\lastreadfilename}%
   \immediate\openout##1=\lastreadfilename\relax}}%
\def\warnopenout#1#2{%
 \readfilename#2\relax
 \def\t@mp{TrashMe,psbjoin.aux,psbjoint.tex,}\uncatcode\t@mp
 \if\matchexpin{\t@mp}{\lastreadfilename,}%
 \else
   \immediate\openin\pst@mpin=\lastreadfilename\relax
   \ifeof\pst@mpin
     \else
     \errhelp{If the content of this file is so precious to you, abort (ie
press x or e) and rename it before retrying.}%
     \errmessage{I'm just about to replace your file named \lastreadfilename}%
   \fi
   \immediate\closein\pst@mpin
 \fi
 \message{\lastreadfilename}%
 \immediate\openout#1=\lastreadfilename\relax}%
{\catcode`\%=12\catcode`\*=14
\gdef\splitfile#1{*
 \readfilename#1\relax
 \immediate\openin\j@insplitin=\lastreadfilename\relax
 \ifeof\j@insplitin
   \message{! I couldn't find and split \lastreadfilename!}*
 \else
   \immediate\openout\j@insplitout=TrashMe
   \message{< Splitting \lastreadfilename\space into}*
   \loop
     \ifeof\j@insplitin
       \immediate\closein\j@insplitin\n@teoffalse
     \else
       \n@teoftrue
       \executeinspecs{\global\read\j@insplitin to\spl@tinline\expandafter
         \ch@ckbeginnewfile\spl@tinline
       \ifc@ntrolline
       \else
         \toks0=\expandafter{\spl@tinline}*
         \immediate\write\j@insplitout{\the\toks0}*
       \fi
     \fi
   \ifn@teof\repeat
   \immediate\closeout\j@insplitout
 \fi\message{>}*
}*
\gdef\ch@ckbeginnewfile#1
 \def\t@mp{#1}*
 \ifx\@mpty\t@mp
   \def\t@mp{#3}*
   \ifx\@mpty\t@mp
     \global\c@ntrollinefalse
   \else
     \immediate\closeout\j@insplitout
     \warnopenout\j@insplitout{#2}*
     \global\c@ntrollinetrue
   \fi
 \else
   \global\c@ntrollinefalse
 \fi}*
\gdef\joinfiles#1\into#2{*
 \message{< Joining following files into}*
 \warnopenout\j@insplitout{#2}*
 \message{:}*
 {*
 \edef\w@##1{\immediate\write\j@insplitout{##1}}*
\w@{
\w@{
\w@{
\w@{
\w@{
\w@{
\w@{
\w@{
\w@{
\w@{
\w@{\string\input\space psbox.tex}*
\w@{\string\splitfile{\string\jobname}}*
\w@{\string\let\string\autojoin=\string\relax}*
}*
 \expandafter\tre@tfilelist#1, \endtre@t
 \immediate\closeout\j@insplitout
 \message{>}*
}*
\gdef\tre@tfilelist#1, #2\endtre@t{*
 \readfilename#1\relax
 \ifx\@mpty\lastreadfilename
 \else
   \immediate\openin\j@insplitin=\lastreadfilename\relax
   \ifeof\j@insplitin
     \errmessage{I couldn't find file \lastreadfilename}*
   \else
     \message{\lastreadfilename}*
     \immediate\write\j@insplitout{
     \executeinspecs{\global\read\j@insplitin to\oldj@ininline}*
     \loop
       \ifeof\j@insplitin\immediate\closein\j@insplitin\n@teoffalse
       \else\n@teoftrue
         \executeinspecs{\global\read\j@insplitin to\j@ininline}*
         \toks0=\expandafter{\oldj@ininline}*
         \let\oldj@ininline=\j@ininline
         \immediate\write\j@insplitout{\the\toks0}*
       \fi
     \ifn@teof
     \repeat
   \immediate\closein\j@insplitin
   \fi
   \tre@tfilelist#2, \endtre@t
 \fi}*
}%
\def\autojoin{%
 \immediate\write\psbj@inaux{\string\into{psbjoint.tex}}%
 \immediate\closeout\psbj@inaux
 \expandafter\joinfiles\GlobalInputList\into{psbjoint.tex}%
}%
%
%
%
\def\centinsert#1{\midinsert\line{\hss#1\hss}\endinsert}%
\def\psannotate#1#2{\vbox{%
  \def\ps@nnotation{#2\global\let\ps@nnotation=\relax}#1}}%
\def\pscaption#1#2{\vbox{%
   \setbox\drawingBox=#1
   \copy\drawingBox
   \vskip\baselineskip
   \vbox{\hsize=\wd\drawingBox\setbox0=\hbox{#2}%
     \ifdim\wd0>\hsize
       \noindent\unhbox0\tolerance=5000
    \else\centerline{\box0}%
    \fi
}}}%
%
\def\at(#1;#2)#3{\setbox0=\hbox{#3}\ht0=0pt\dp0=0pt
  \rlap{\kern#1\vbox to0pt{\kern-#2\box0\vss}}}%
%
\newdimen\gridht \newdimen\gridwd
\def\gridfill(#1;#2){%
  \setbox0=\hbox to 1\pscm
  {\vrule height1\pscm width.4pt\leaders\hrule\hfill}%
  \gridht=#1
  \divide\gridht by \ht0
  \multiply\gridht by \ht0
  \gridwd=#2
  \divide\gridwd by \wd0
  \multiply\gridwd by \wd0
  \advance \gridwd by \wd0
  \vbox to \gridht{\leaders\hbox to\gridwd{\leaders\box0\hfill}\vfill}}%
%
\def\fillinggrid{\at(0cm;0cm){\vbox{%
  \gridfill(\drawinght;\drawingwd)}}}%
%
%
\def\textleftof#1:{%
  \setbox1=#1
  \setbox0=\vbox\bgroup
    \advance\hsize by -\wd1 \advance\hsize by -2em}%
\def\textrightof#1:{%
  \setbox0=#1
  \setbox1=\vbox\bgroup
    \advance\hsize by -\wd0 \advance\hsize by -2em}%
\def\endtext{%
  \egroup
  \hbox to \hsize{\valign{\vfil##\vfil\cr%
\box0\cr%
\noalign{\hss}\box1\cr}}}%
%
\def\frameit#1#2#3{\hbox{\vrule width#1\vbox{%
  \hrule height#1\vskip#2\hbox{\hskip#2\vbox{#3}\hskip#2}%
        \vskip#2\hrule height#1}\vrule width#1}}%
\def\boxit#1{\frameit{0.4pt}{0pt}{#1}}%
\catcode`\@=12 
%
 \psfordvips   

\begin{figure}
$$\psannotate{\psboxscaled{600}{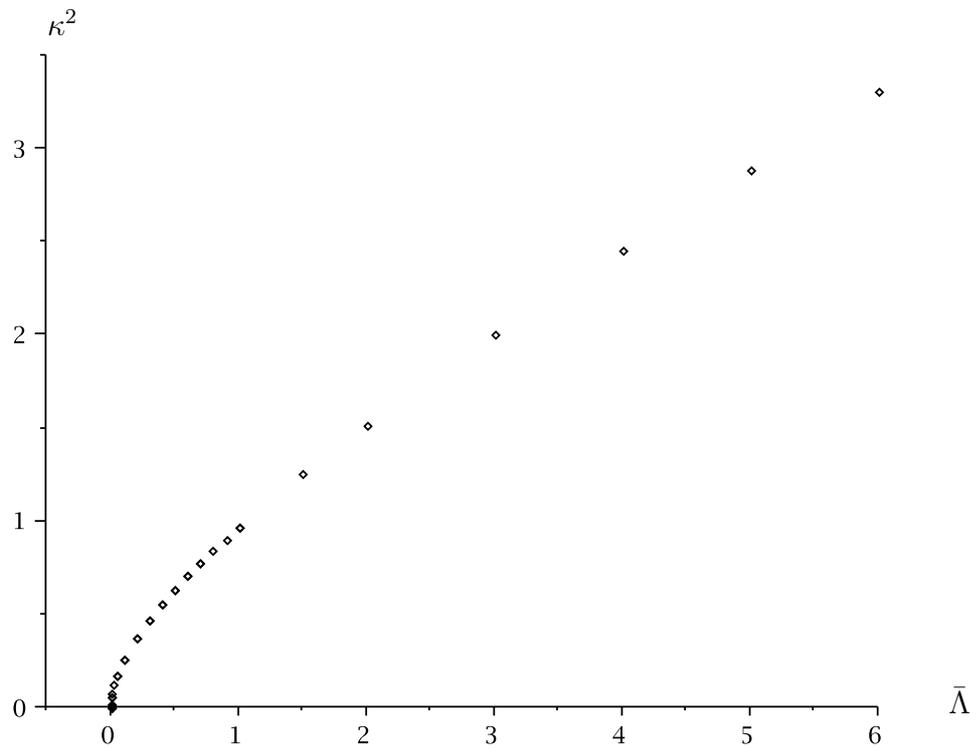}}{
\at(21\pscm;1\pscm){$\bar \Lambda $} \at(1\pscm;16\pscm){$\kappa^2$}}$$
\caption{The fine-tuned values of $\kappa $ and $\bar \Lambda$, for which a
compacton exists, in the 
$\bar \Lambda$-$\kappa^2$ plane, for some selected values. }
\end{figure}

\begin{figure}
$$\psannotate{\psboxscaled{700}{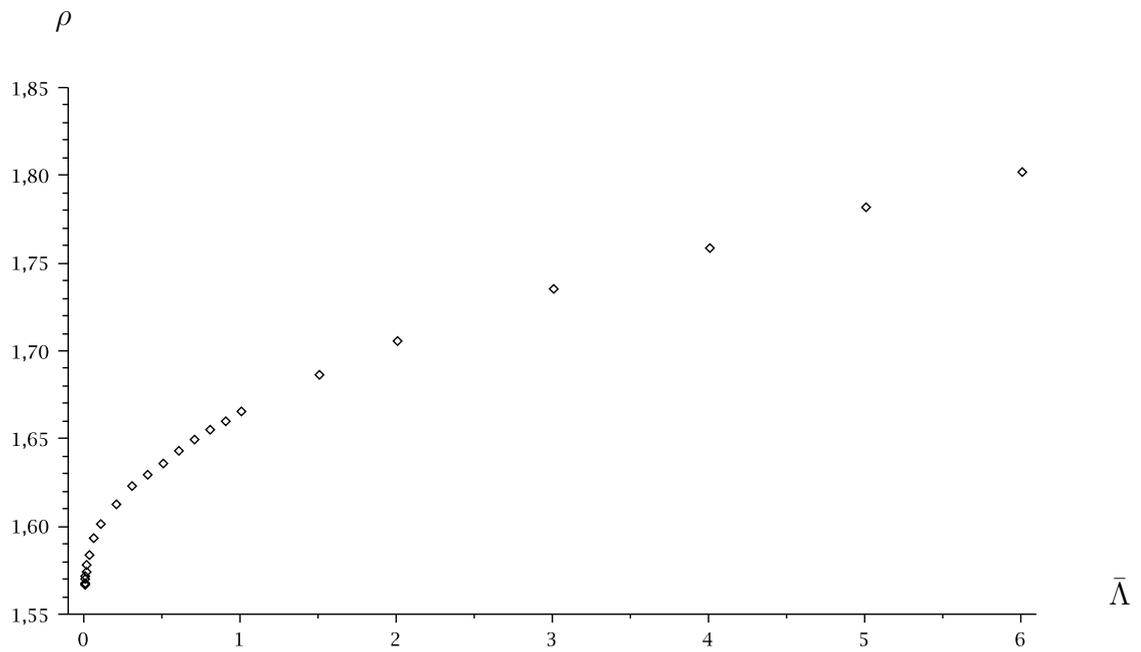}}{
\at(21\pscm;1\pscm){$\bar \Lambda $} \at(1\pscm;12\pscm){$\rho$}}$$
\caption{The compacton radius $\rho$ for some selected
values of $\bar \Lambda$
(and for the corresponding fine-tuned values of $\kappa^2$ such that the
compacton exists), in the 
$\bar \Lambda$-$\rho$ plane. }
\end{figure}

\begin{figure}
$$\psannotate{\psboxscaled{600}{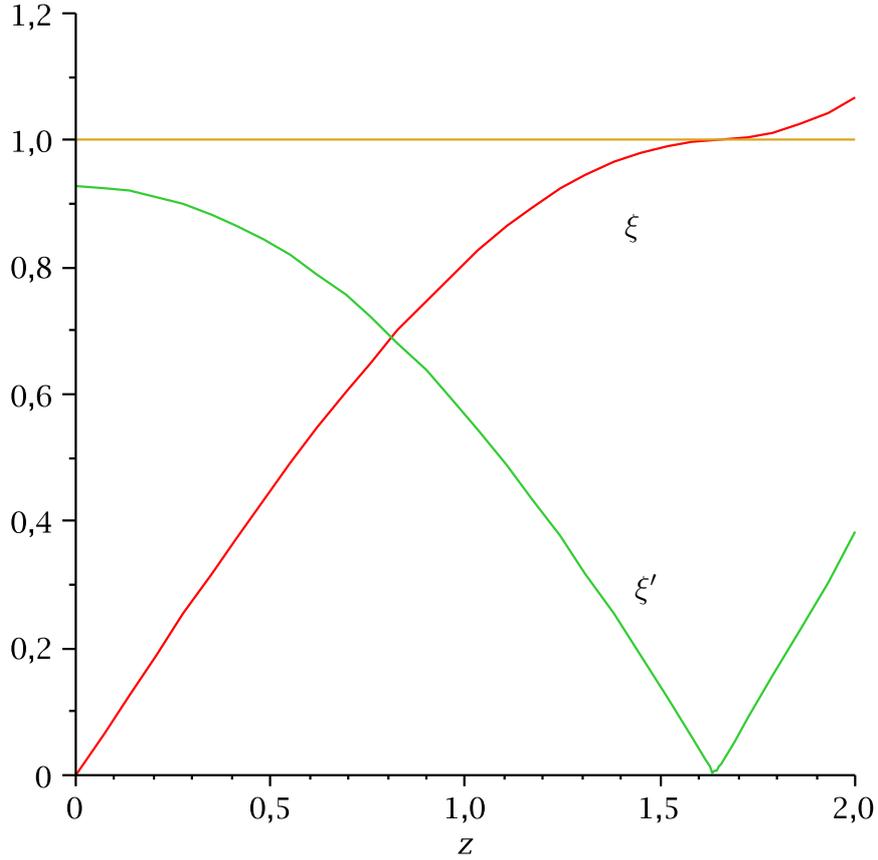}}{
\at(14\pscm;14\pscm){$\xi $} \at(14\pscm;6\pscm){$\xi '$}}$$
\caption{Shooting from the center: for 
$\bar \Lambda =0.5 $ and for the fine-tuned value $ \kappa^2 = 0.633226 $
the functions $\xi$ and $\xi ' \equiv \xi_z$ are shown. The compacton field
$\xi$ takes its vacuum value $\xi =1$ exactly at the point where  $\xi_z =0$
(the compacton boundary). To the right of the compacton boundary the numerical
integration selects the compacton configuration instead of the (correct) vacuum
configuration, because numerically $\xi_z$ never is exactly zero (because in
that case the numerical integration would break down). 
}
\end{figure}

\begin{figure}
$$\psannotate{\psboxscaled{600}{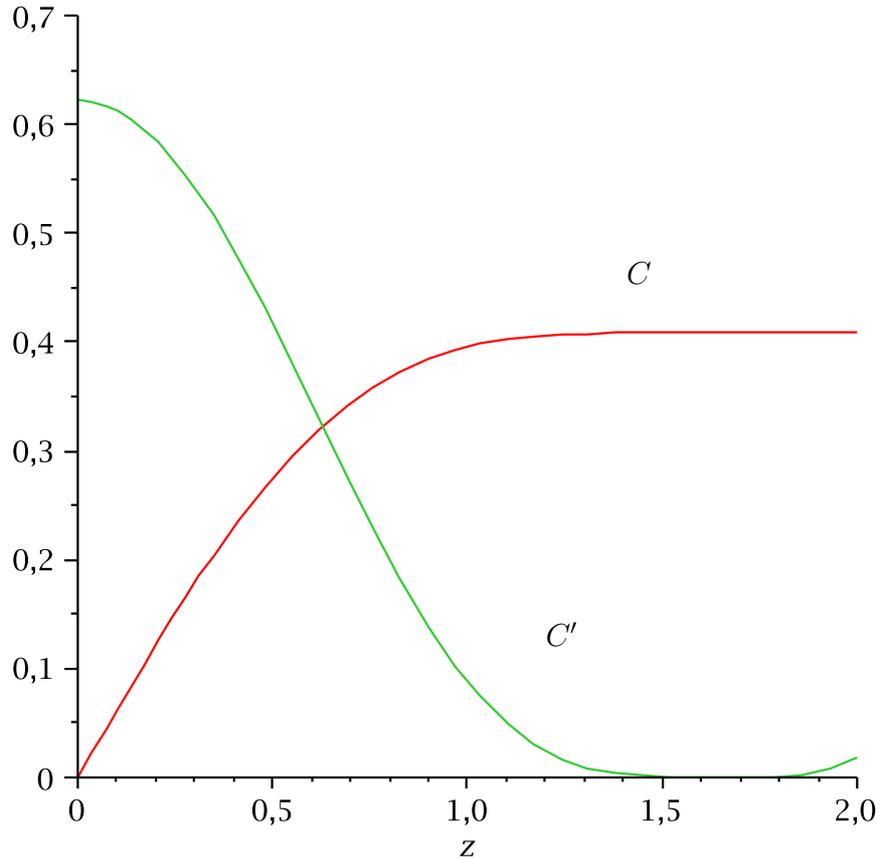}}{
\at(14\pscm;13\pscm){$C $} \at(12\pscm;5\pscm){$C '$}}$$
\caption{Shooting from the center: for 
$\bar \Lambda =0.5 $ and for the fine-tuned value $ \kappa^2 = 0.633226 $
(the same values as in Figure 3)
the functions $C$ and $C ' \equiv C_z$ are shown. The field $C$
takes its vacuum value $C =\sqrt{\bar \Lambda /3}$ exactly at the 
point where  $\xi_z =C_z =0$
(the compacton boundary). The curve is very flat near the compacton boundary,
because the first four derivatives of $C$ at the boundary are zero.
To the right of the compacton boundary the numerical
integration selects the compacton configuration instead of the (correct) vacuum
configuration, for the same reason as in Figure 3. 
}
\end{figure}

\begin{figure}
$$\psannotate{\psboxscaled{600}{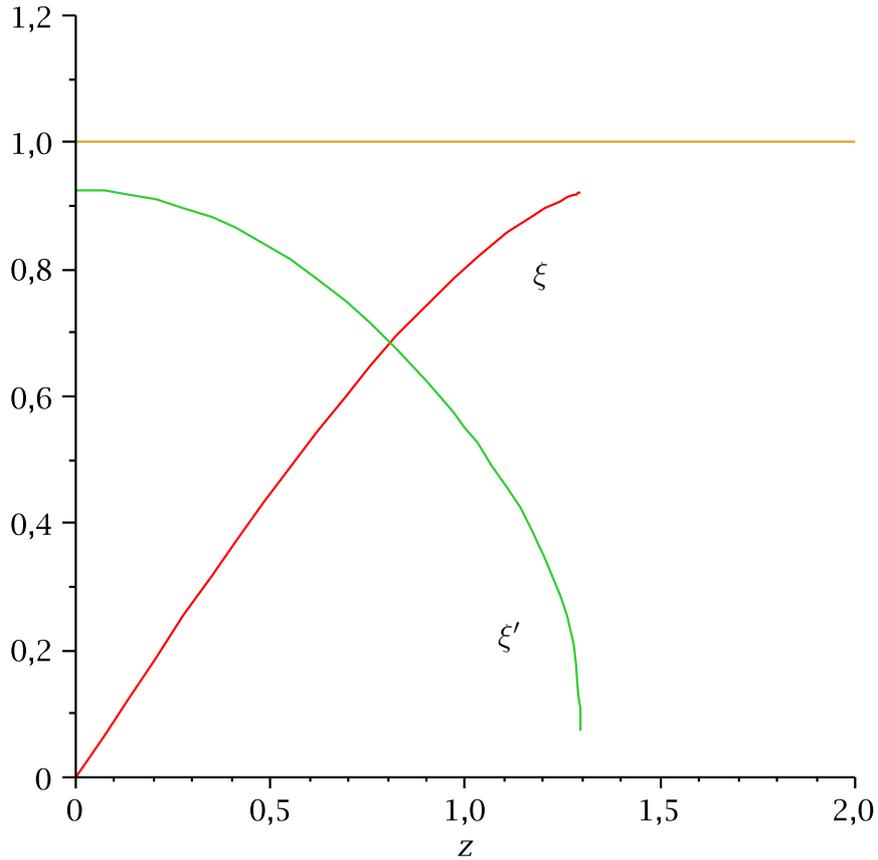}}{
\at(12\pscm;13\pscm){$\xi $} \at(11\pscm;5\pscm){$\xi '$}}$$
\caption{Shooting from the center: for 
$\bar \Lambda =0.5 $ and for $ \kappa^2 = 0.621335 $ (slightly smaller than the
fine-tuned value)
the functions $\xi$ and $\xi ' \equiv \xi_z$ are shown. The field $\xi '$
reaches the value $\xi ' =0$ before $\xi $ reaches its vacuum value 
$\xi =1$. Further, all higher derivatives of $\xi$ become singular at that
point. Therefore, the numerical integration breaks down at this point. 
The numerical value of the point where the integration breaks down is
$z=1.29653$.
}
\end{figure}

\begin{figure}
$$\psannotate{\psboxscaled{600}{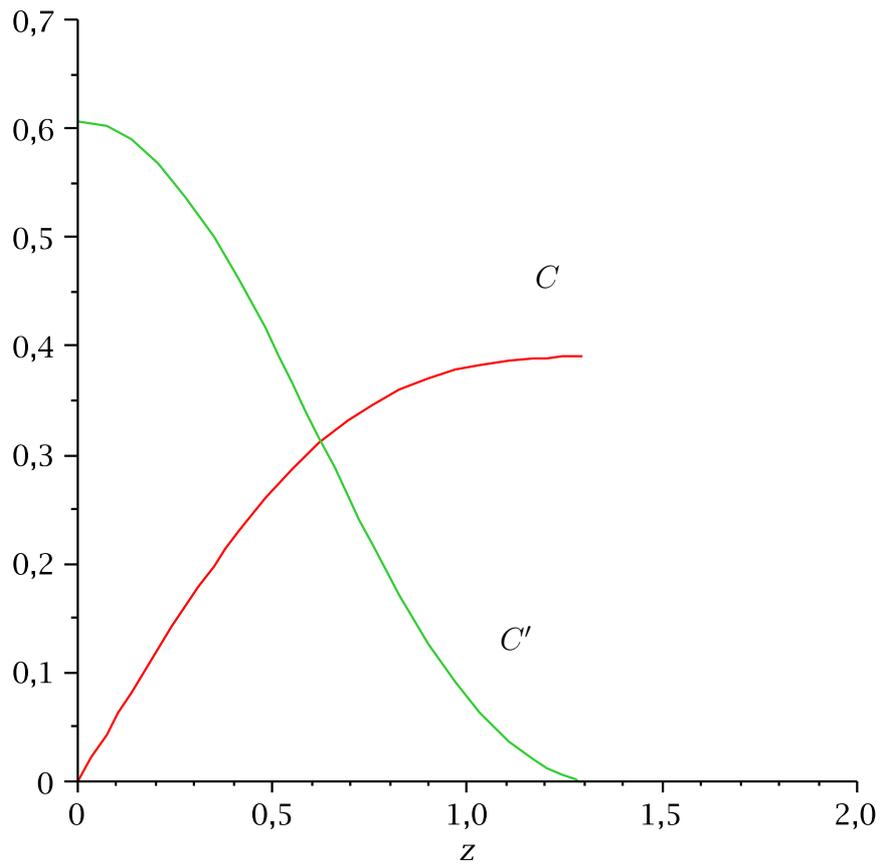}}{
\at(12\pscm;13\pscm){$C $} \at(11\pscm;5\pscm){$C '$}}$$
\caption{Shooting from the center: for 
$\bar \Lambda =0.5 $ and $ \kappa^2 = 0.621335 $ (slightly smaller than the
fine-tuned value), like in Figure 4,
the functions $C$ and $C ' \equiv C_z$ are shown. The field $C '$
reaches the value $C ' =0$ before $\xi $ and $C$ reach their vacuum values. 
The numerical integration breaks down at this point because of the singularity
in the integration for $\xi$, like in Figure 5. 
}
\end{figure}

\begin{figure}
$$\psannotate{\psboxscaled{600}{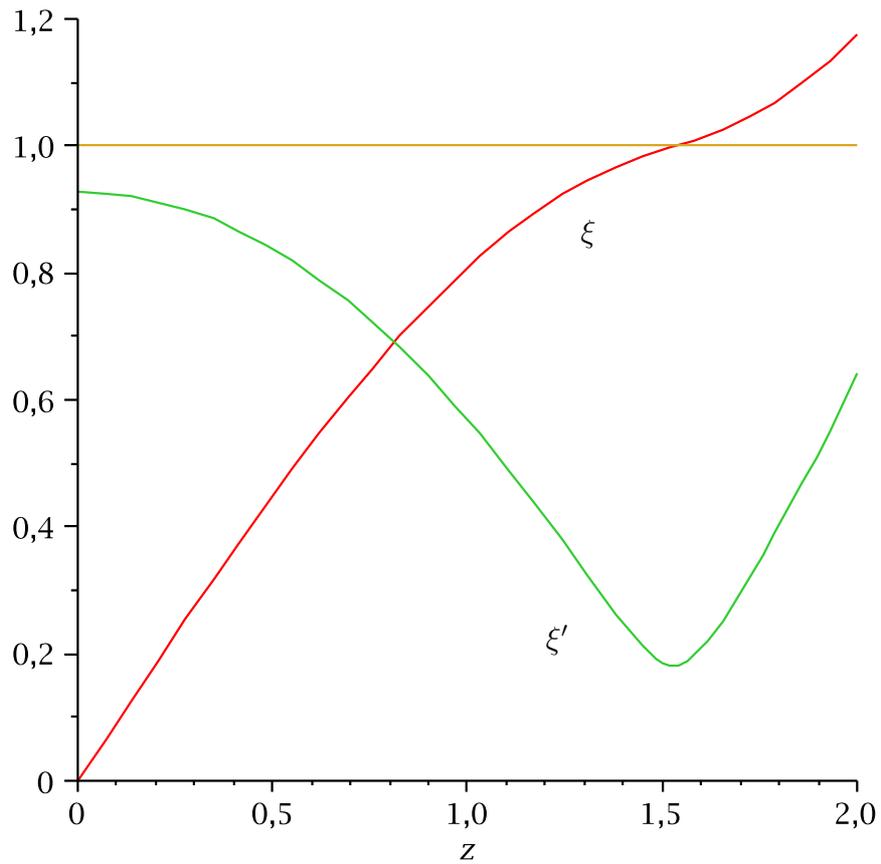}}{
\at(13\pscm;14\pscm){$\xi $} \at(12\pscm;5\pscm){$\xi '$}}$$
\caption{Shooting from the center: for 
$\bar \Lambda =0.5 $ and for $ \kappa^2 = 0.633693 $ (slightly bigger than the
fine-tuned value)
the functions $\xi$ and $\xi ' \equiv \xi_z$ are shown. The field $\xi '$
never reaches the value $\xi ' =0$, therefore the field $\xi$ never settles
down at its vacuum value.
}
\end{figure}

\begin{figure}
$$\psannotate{\psboxscaled{600}{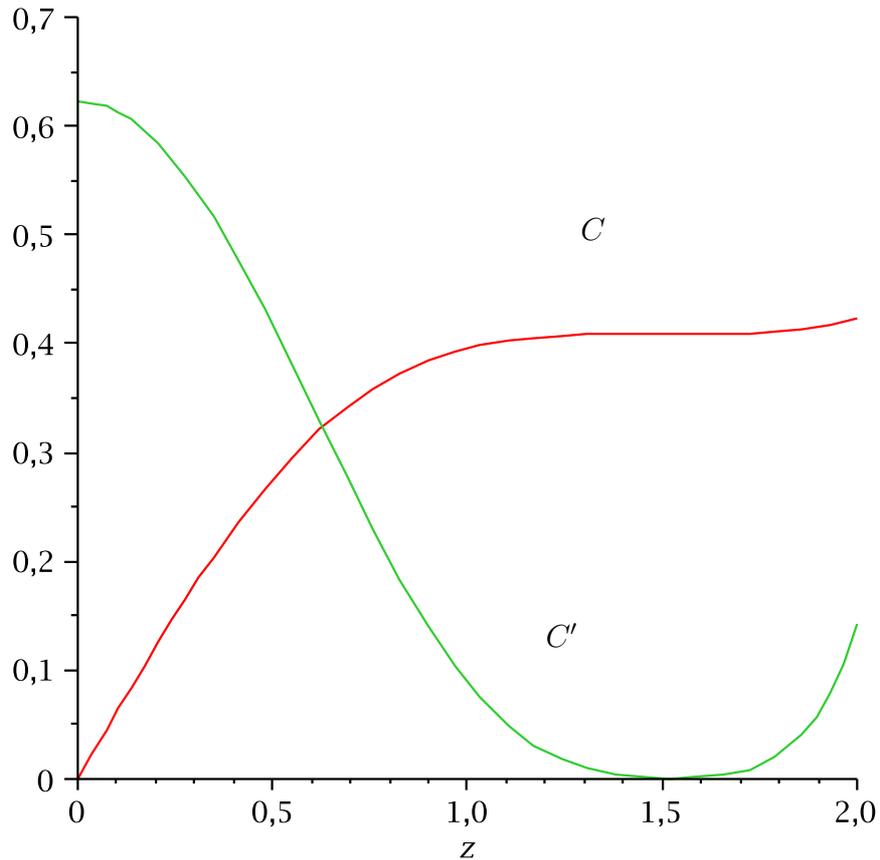}}{
\at(13\pscm;14\pscm){$C $} \at(12\pscm;5\pscm){$C '$}}$$
\caption{Shooting from the center: for 
$\bar \Lambda =0.5 $ and for $ \kappa^2 = 0.633693 $ (slightly bigger than the
fine-tuned value), like in Figure 7,
the functions $C$ and $C ' \equiv C_z$ are shown. The field $C '$
never reaches the value $C ' =0$, therefore the field $C$ never settles
down at its vacuum value (the fact that $C'$ does not reach zero is not
obvious from the figure, because it takes very small values numerically near
the region where $\xi \sim 1$. But it may be checked easily by direct
numerical calculation, or by amplifying that region in the figure). 
}
\end{figure}

\begin{figure}
$$\psannotate{\psboxscaled{500}{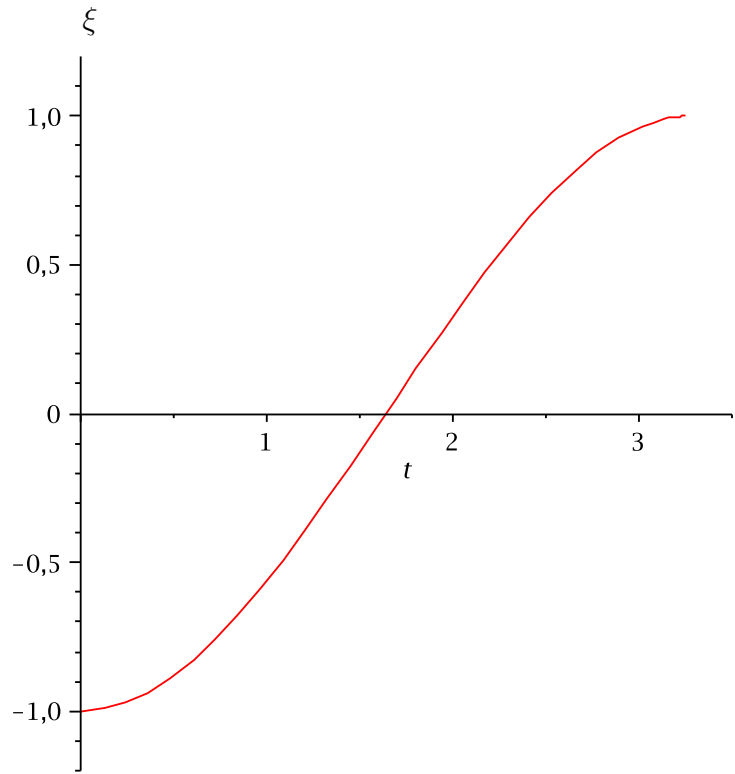}}{
\at(2.5\pscm;20\pscm){$\xi $} }$$
\caption{Shooting from the lower boundary: for 
$\bar \Lambda =0.5 $ and for the fine-tuned value $ \kappa^2 = 0.633226 $,
like in Figure 3,
the function $\xi$ is shown. The compacton field
$\xi$ takes its opposite vacuum value $\xi =+1$ exactly at the point where  
$\xi_z =0$ again
(the upper compacton boundary). The graph of the function $\xi$ is exactly odd
about the compacton center (where $\xi =0$). This compacton center $t=z_0$ 
 is at $z_0 =1.6414529$ 
}
\end{figure}

\begin{figure}
$$\psannotate{\psboxscaled{500}{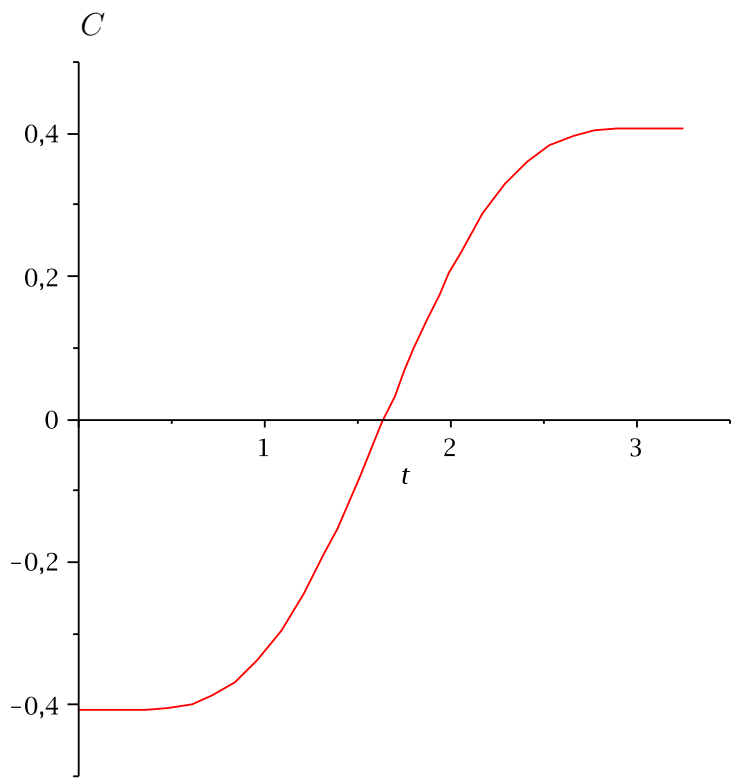}}{
\at(2.5\pscm;20\pscm){$C $} }$$
\caption{Shooting from the lower boundary: for 
$\bar \Lambda =0.5 $ and for the fine-tuned value $ \kappa^2 = 0.633226 $
(the same values as in Figure 3)
the function $C$ is shown. The field $C$
takes its opposite vacuum value $C =+\sqrt{\bar \Lambda /3}$ exactly at the 
point where  $\xi_z =C_z =0$ again
(the upper compacton boundary). 
The curve is very flat near the compacton boundaries,
because the first four derivatives of $C$ at the boundaries are zero.
The graph of the function $C$ is exactly odd
about the compacton center (where $\xi =C=0$). Numerically, this center is at
$t=1.6414542$, which agrees with the compacton center for $\xi$
in Figure 9 in the first
six digits. 
}
\end{figure}

\begin{figure}
$$\psannotate{\psboxscaled{500}{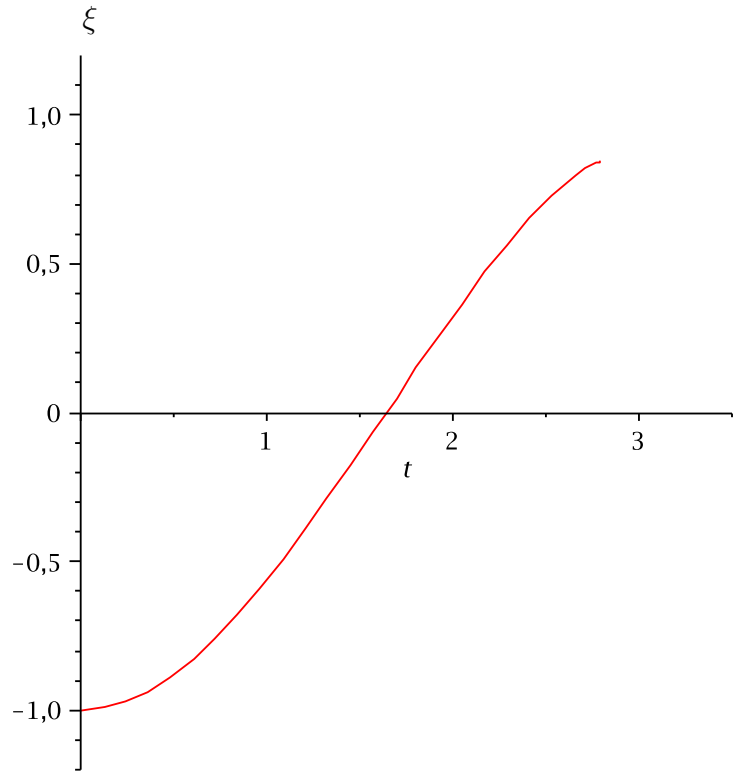}}{
\at(2.5\pscm;20\pscm){$\xi $} }$$
\caption{Shooting from the lower boundary: for 
$\bar \Lambda =0.5 $ and for $ \kappa^2 = 0.608400 $ (slightly smaller than the
fine-tuned value), 
the function $\xi$ is shown. The field $\xi '$
reaches the value $\xi ' =0$ before $\xi $ reaches its vacuum value 
$\xi =1$. Therefore, the numerical integration breaks down at this point. 
The numerical value of the point where the integration breaks down is
$t=2.79688$. The numerical value of the ``compacton center'' 
where $\xi (t)=0$ is $t=1.64236$.
}
\end{figure}

\begin{figure}
$$\psannotate{\psboxscaled{500}{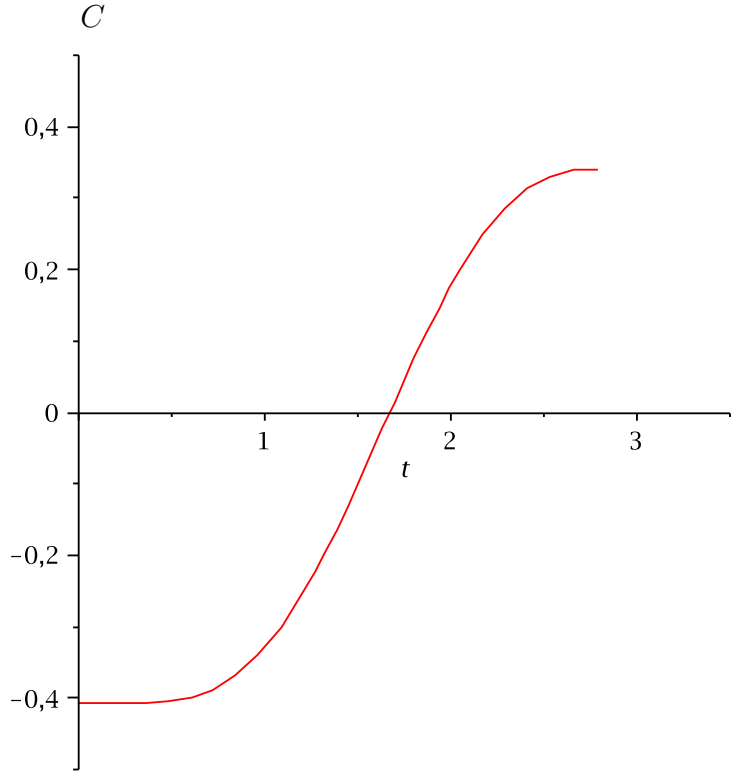}}{
\at(2.5\pscm;20\pscm){$C $} }$$
\caption{Shooting from the lower boundary: for 
$\bar \Lambda =0.5 $ and $ \kappa^2 = 0.608400 $ (slightly smaller than the
fine-tuned value), 
the function $C$ is shown. The field $C '$
reaches the value $C ' =0$ before $\xi $ and $C$ reach their vacuum values. 
The numerical integration breaks down at this point because of the singularity
in the integration for $\xi$, like in Figure 11. The field $C(t)$ takes the
value zero at $t=1.67201$, which is markedly different from the point where
the field $\xi$ takes the value zero in Figure 11.
}
\end{figure}

\begin{figure}
$$\psannotate{\psboxscaled{500}{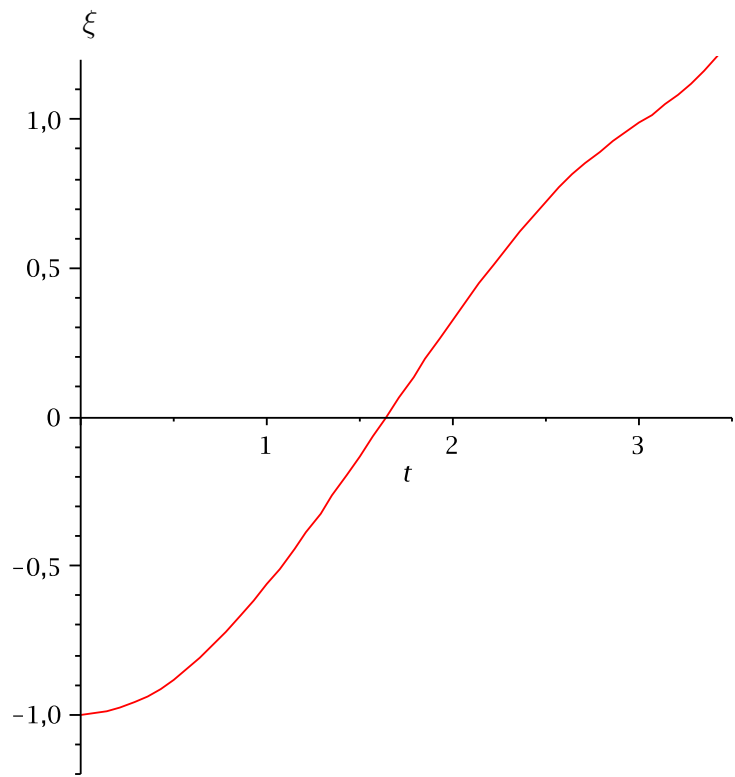}}{
\at(2.5\pscm;20\pscm){$\xi $} }$$
\caption{Shooting from the lower boundary: for 
$\bar \Lambda =0.5 $ and for $ \kappa^2 = 0.640000 $ (slightly bigger than the
fine-tuned value), 
the function $\xi$ is shown. The field $\xi '$
never reaches the value $\xi ' =0$, therefore the field $\xi$ never settles
down at its vacuum value.
The point where $\xi (t)=0$ is at $t=1.641214$.
}
\end{figure}

\begin{figure}
$$\psannotate{\psboxscaled{500}{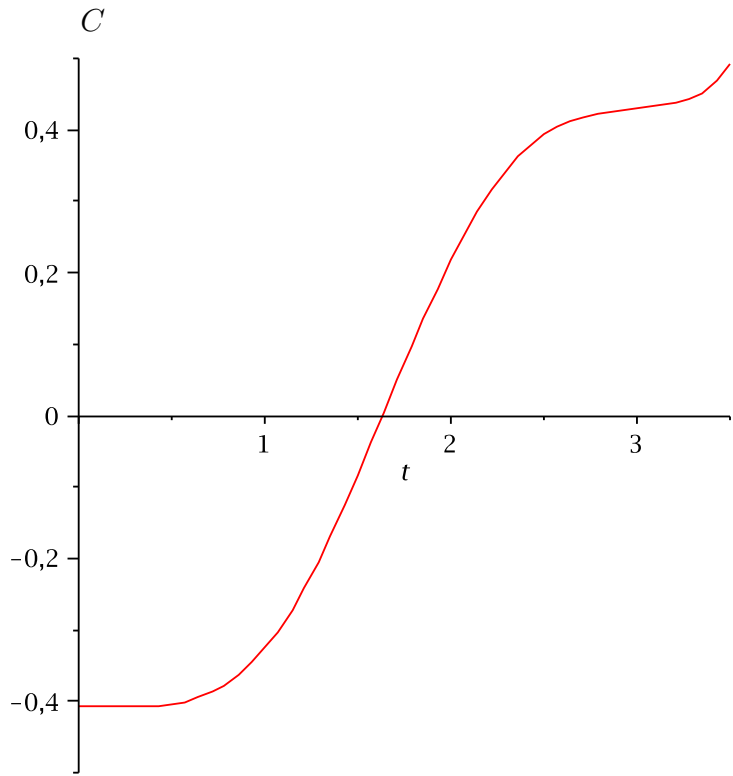}}{
\at(2.5\pscm;20\pscm){$C $} }$$
\caption{Shooting from the lower boundary: for 
$\bar \Lambda =0.5 $ and for $ \kappa^2 = 0.640000 $ (slightly bigger than the
fine-tuned value), 
the function $C$ is shown. The field $C '$
never reaches the value $C ' =0$, therefore the field $C$ never settles
down at its vacuum value. 
The field $C(t)$ takes the
value zero at $t=1.63363$, which is markedly different from the point where
the field $\xi$ takes the value zero in Figure 13.
}
\end{figure}

\begin{figure}
$$\psannotate{\psboxscaled{700}{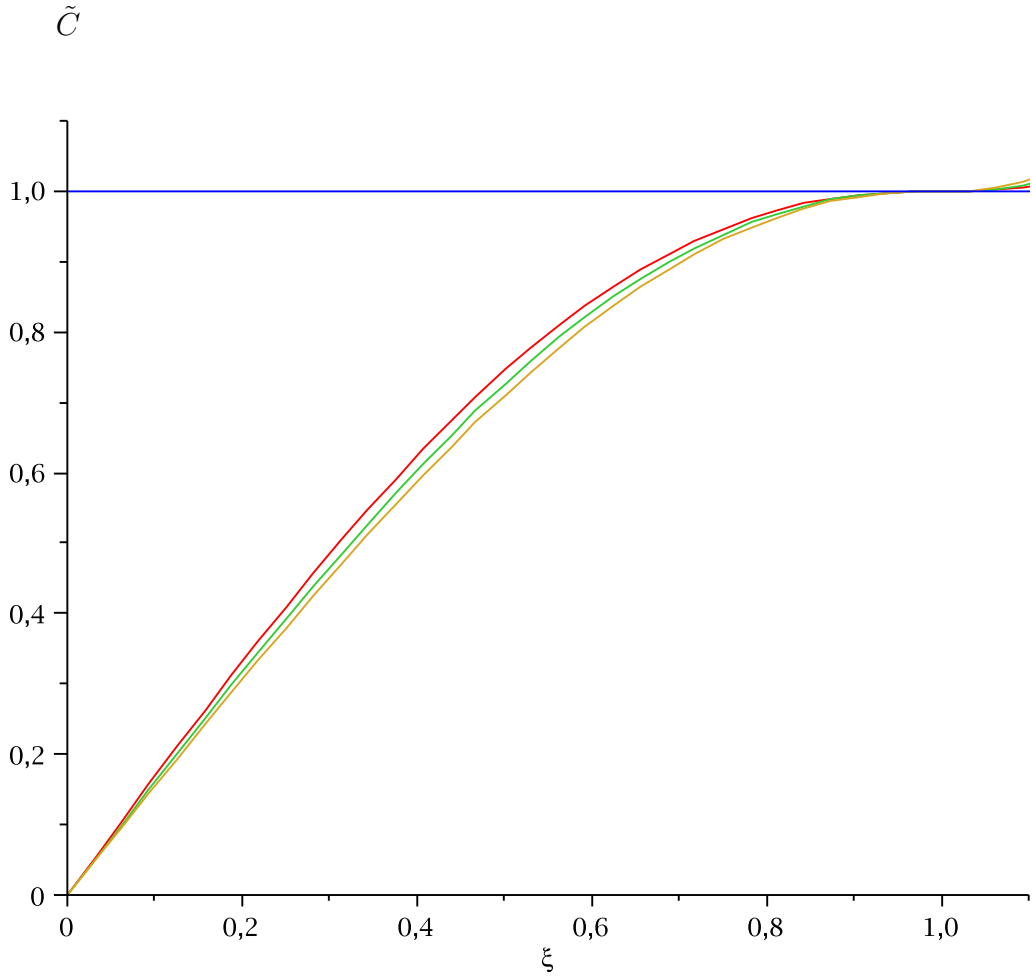}}{
\at(1.3\pscm;18\pscm){$\tilde C $} }$$
\caption{Integration of the orbit equation (\ref{orbit-eq}) for some
fine-tuned values of $\kappa$ and $\bar \Lambda$. The values of the
parameters and the resulting $\tilde C(\xi =1)$ are \newline
upper curve: $\bar \Lambda = 0.005\, ,\quad  \kappa = 0.2303457790 \,  , \quad
\tilde C (\xi =1)= 1.0000029 $, \newline
middle curve: $\bar \Lambda = 1.5\, ,\quad  \kappa = 1.119251347 \,  ,\quad
\tilde C (\xi =1)= 1.0000074 $, \newline
lower curve: $\bar \Lambda = 6.0\, ,\quad  \kappa = 1.818434381 \, , ,\quad
\tilde C (\xi =1)= 0.9999953 $. \newline
So $\tilde C$ reaches the compacton value $\tilde C (\xi =1)=1$ 
to a high precision. }
\end{figure}

\begin{figure}
$$
\psannotate{
\psboxscaled{300}{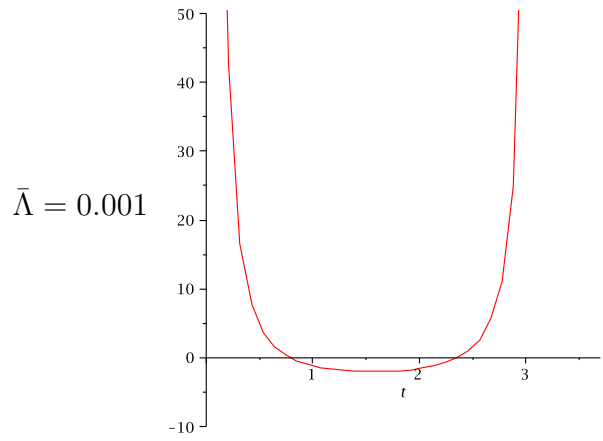}
}{
\at(-6\pscm;10\pscm){$\bar \Lambda = 0.001 $} }
$$

$$
\psannotate{
\psboxscaled{300}{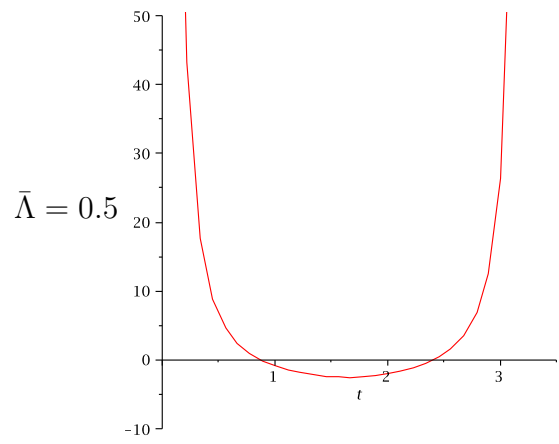}
}{
\at(-4\pscm;10\pscm){$\bar \Lambda = 0.5 $} }
$$

$$
\psannotate{
\psboxscaled{300}{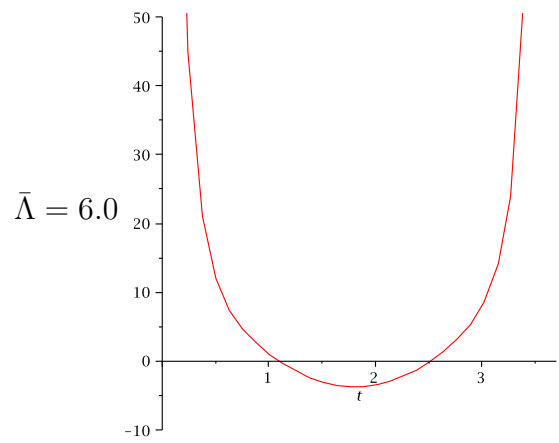}
}{
\at(-4\pscm;10\pscm){$\bar \Lambda = 6.0 $} }
$$
\caption{The potential $U$ for the effective Schroedinger equation of
Section 6, for the values $\bar \Lambda =0.001, 0.5, 6.0$. }
\end{figure}

\newpage

\begin{figure}
$$
\psannotate{
\psboxscaled{500}{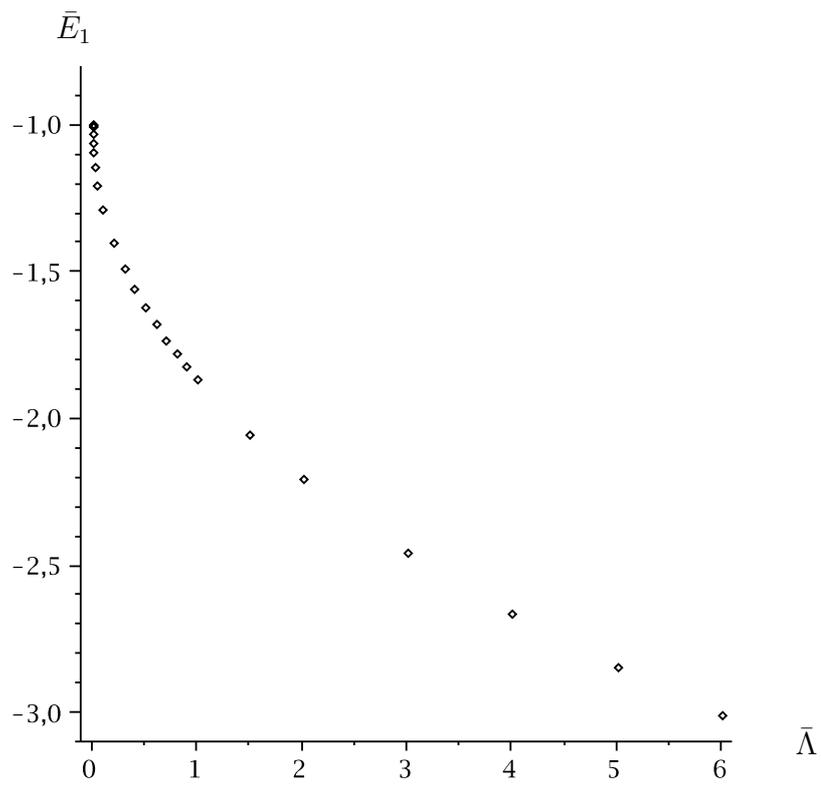}
}{
\at(1.8\pscm;20\pscm){$\bar E_1$} \at(21\pscm;1\pscm){$\bar \Lambda $}}
$$
\caption{The first energy eigenvalue of the infinite square wall potential
of Section 6. The eigenvalue is negative for all values of $\bar \Lambda$ }
\end{figure}

\begin{figure}
$$
\psannotate{
\psboxscaled{500}{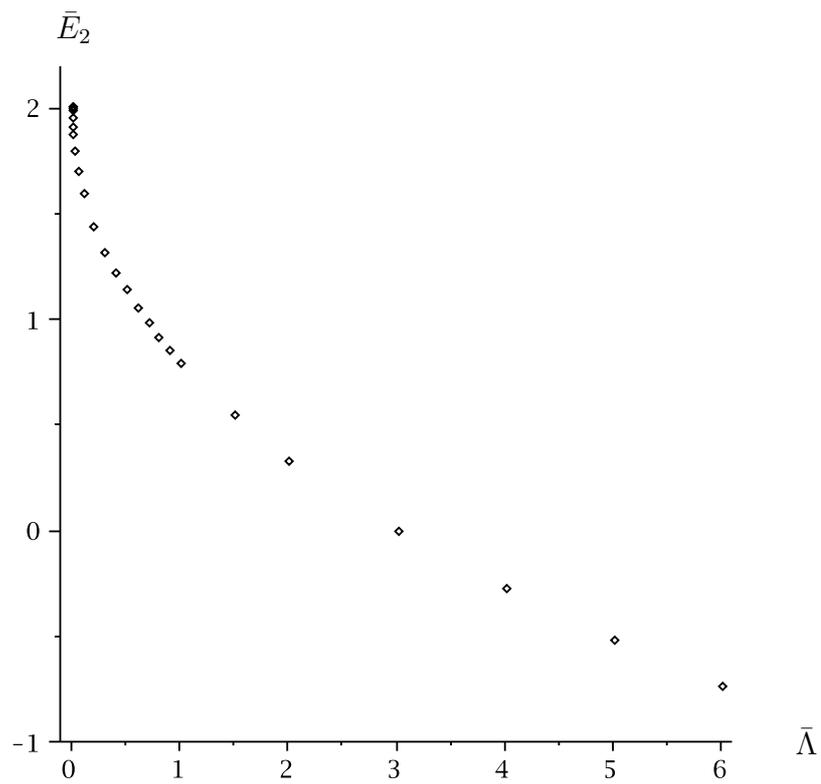}
}{
\at(1.8\pscm;20\pscm){$\bar E_2$} \at(21\pscm;1\pscm){$\bar \Lambda $}}
$$
\caption{The second energy eigenvalue of the infinite square wall potential
of Section 6. The eigenvalue is positive for sufficiently small values 
of $\bar \Lambda$ (at least till $\bar \Lambda =2$) }
\end{figure}

\end{document}